\documentclass[aps,pra,twocolumn,showpacs,floatfix,amsmath,amssymb,longbibliography,10pt]{revtex4-1}

\usepackage{graphicx}
\usepackage{dcolumn}
\usepackage{bm}
\usepackage{amssymb}
\usepackage{amsmath}
\usepackage{float}
\usepackage{multirow}
\usepackage{tgtermes}
\usepackage{ulem}
\usepackage{color}
\usepackage[english]{babel}
\usepackage{txfonts}
\usepackage[altg]{qtxmath}
\usepackage{microtype}
\usepackage{slashed}
\usepackage{diagbox}

\makeatletter
\newcommand*{\rom}[1]{\expandafter\@slowromancap\romannumeral #1@}
\makeatother

\begin{document}

\author{L. Aliani}
\author{Q.~Z. Lv}
\email{qingzheng.lyu@mpi-hd.mpg.de}
\affiliation{Max-Planck-Institut f\"{u}r Kernphysik, Saupfercheckweg 1,  69117 Heidelberg, Germany }

\date{\today}

\title{Influence of the rotational sense of two colliding laser beams on \\the radiation of an ultrarelativistic electron}

\begin{abstract}

With analytical treatment, the classical dynamics of an ultrarelativistic electron in two counter-propagating circularly polarized strong laser beams with either co-rotating or counter-rotating direction are considered. Assuming that the particle energy is the dominant scale in the setup, an approximate solution is derived and the influence of the rotational sense on the dynamics is analyzed. Qualitative differences in both electron energy and momentum are found for the laser beams being co-rotating or counter-rotating and are confirmed by the exact numerical solution of the classical equation of motion. Despite of these differences in the electron trajectory, the radiation spectra of the electron do not deviate qualitatively from each other for configurations with varying rotational directions of the laser beams. Here, the radiation of an ultrarelativistic electron interacting with counterpropagating laser beams is given in the framework of the Baier-Katkov semi-classical approximation. Several parameter regimes are considered and the spectra resulting from the two scenarios all have the same shape and only differ quantitatively by a few percent. 

\end{abstract}

\maketitle

\section{introduction}
	\label{sec:intro}

Ultrastrong laser techniques, presently available \cite{Vulcan1,Yoon_2019} and near future facilities \cite{ELI,XCELS}, will allow investigation of extreme regimes of nonlinear Quantum Electrodynamics (QED) processes. $\gamma$-photons can be produced by electron radiation, which may further generate electron-positron pairs, and even avalanches of pairs through cascade processes \cite{Marklund_2006,RMP_2012,Heinzl_2012,Dunne2014,Turcu_2019}. Multiple photon emissions and large quantum recoils of emitted photons result in the emergence of conspicuous classical and quantum radiation reaction effects \cite{Cole_2018,Poder_2018}. In the realm of the standard QED, these nonlinear  processes correspond to including a huge number of orders in the perturbative procedure, which yields numerous Feynman diagrams and renders the calculation intractable. In order to overcome this obstacle, the strong-field QED has been introduced \cite{Furry_1951}, which replaces the free particle wave function in the calculations by the wave function in the presence of the strong field under consideration. Since the strong field is regarded as a classical field here, the transition from the standard to the strong-field QED is determined by the dimensionless field strength $\xi \equiv -ea/m$ where $e,m$ are the electron charge and mass respectively while $a=\sqrt{-A^2}$ with $A_{\mu}$ being the electromagnetic four vector potential. Relativistic units $\hbar=c=1$ are used throughout the paper, unless specified otherwise.

Unfortunately, the exact analytical wave functions for an electron in classical background fields exist only for a few special field configurations \cite{Bagrov_1990}, which increases the difficulty of therotically studying the strong field QED processes. For a general configuration, several approximate approaches are developed to investigate the nonlinear QED processes. For very high field intensities ($\xi\gg 1$), the most common way is to approximate the emission or pair creation probability to that in the presence of a locally constant field. However, recently deficiencies and breakdowns of the local constant field approximation (LCFA) have been observed in various regimes \cite{DiPiazza_2018,DiPiazza_2019,Ilderton_2019a,Ilderton_2019b,Podszus_2019,Lv_2021a}. Another more accurate approach is the so-called Wentzel-Kramers-Brillouin (WKB) approximation \cite{Popov_1997,Gersten_1975,Mocken_2010,DiPiazza_2014,DiPiazza_2015,DiPiazza_2016,DiPiazza_2017}. This approximation is closely connected to the classical description and can only be obtained in cases in which the electron's classical equation of motion is soluble. Inspired by the spirit of the WKB approximation, Baier and Katkov have developed the semi-classical operator approach, which can express the amplitudes of strong-field QED processes in general background fields as a function of the electron's classical trajectory in the fields \cite{Katkov_1968,Baier_b_1994, Landau_4}.
	
The experimental limits of investigating the strong field QED processes is the field strength obtained in the laboratory. The desire to increase the effective field strength even more with a fixed laser beam energy gave rise to the concept of multi-beam configurations and the notion of a dipole wave \cite{Bulanov_2010_a,Golla_2012,Gonoskov_2012,Bashinov_2013,Bashinov_2019,Magnusson_2019}. One of the simplest cases of multi-beam configuration is the counterpropagating wave (CPW) setup, which is an attractive setup to study strong-field QED effects \cite{Kirk_2009,Bulanov_2010,Gonoskov_2014,Gong_2017,Grismayer_2017} and a favorable configuration for QED cascades \cite{Grismayer_2016,Jirka_2016} or new x-ray sources \cite{Lv_2022}. Especially, the field configuration of a rotating electric field, which minics the antinode of a standing wave, is widely used to study the pair creation mechanism \cite{Brezin_1970,Raicher_2020,Villalba_2019,Dunne_2008}. It is remarkable that in this setup electron trapping dynamics can also be observed, which strongly depend on the nature of the radiation reaction \cite{Gonoskov_2014,Kirk_2016}. The dynamics of a particle in the presence of the CPW configuration are determined by the ratio of the laser frequencies as seen in the electron's average rest frame. If this ratio is close to unity, the system is resonant, giving rise to phenomena such as the Kapitza-Dirac effect \cite{Kapitza_1933,Batelaan_2007,Ahrens_2012,Mueller_2017} and stimulated Compton emission \cite{Friedman_1988,Pantell_1968,Fedorov_1981,Avetissian_b_2016}. The latter is the operating principle for free-electron lasers \cite{Saldin_1995}. On the other hand, the classical and quantum equation of motion for an electron in the non-resonant regime, where the above mentioned ratio is far from one, were also investigated using various approximations \cite{King_2016,Hu_2015,Lv_2021b}.

In this work, the classical dynamics and the radiation processes of an ultrarelativistic electron moving in circularly polarized CPW are considered in the non-resonant regime for both co- and counter-rotating arrangements of the two laser beams. The classical trajectory of the electron is obtained both analytically and numerically. We generalized the approach employed in Ref.~\cite{Lv_2021b}, where the analytical solution is based on an approximation that has imposed a restriction on the laser parameters and the electron's initial momentum, namely, the electron's average energy must be the dominant energy scale in the system ($\xi_1 \xi_2\ll \gamma^2$ with $\xi_1, \xi_2$ being the laser fields' strength and $\gamma$ the average Lorentz factor of the electron in the fields). This solution is verified by a fully numerical calculation. By applying this analytical solution in the Baier-Katkov operator method, we also studied the radiation properties of the electron in the background field. Particularly, the influence of the rotational sense of the two laser beams on the radiation spectra is investigated and shows that though the dynamics are qualitatively different in the co- or counter-rotating case, the radiation spectra of the electron have a similar shape and differ only quantitatively by a few percent.   

The paper is organized as follows. Sec.~\ref{sec:traj} gives the derivations of the electron's classical trajectory in CPW with arbitrary rotating direction, and compares it with a fully numerical solution. The influence of the laser's rotation direction on the electron dynamics is investigated. The calculation of the photon emission matrix elements and the corresponding emission formula are given in Sec.~\ref{sec:radi}. Employing the emission probability formula, we also analyze the effects of the sense of rotation on the radiation spectrum. We conclude our paper in Sec.~\ref{sec:conc} with a discussion of the main findings in this work.

\section{The classical dynamics}
	\label{sec:traj}
	
In this section the mathematical formulation of the CPW problem is introduced. The approximated solution and its validity condition are derived for arbitrary laser rotating directions. The classical equation of motion for an electron in the presence of an electromagnetic field reads
\begin{equation}
	\label{eq:equ.of.mov}
	\frac{d P^{\mu}}{d \tau} = \frac{e}{m} F^{\mu \nu} P_{\nu}\,,
\end{equation}
where $\tau$ is the proper time, $P_{\mu}$ is the particle's four-momentum and $F_{\mu \nu} \equiv \partial_{\mu} A_{\nu}- \partial_{\nu} A_{\mu} $ is the electromagnetic field tensor. The vector potential corresponding to the CPW configuration is $A=A_1+A_2$, where
\begin{eqnarray}
	\label{eq:vec.pot.def}
	\begin{aligned}
		A_1 \equiv a_1 g_1(\phi_1) \left[ \cos \phi_1 e_x + \epsilon_1 \sin \phi_1 e_y \right], \\
		A_2 \equiv a_2 g_2(\phi_2) \left[ \cos \phi_2 e_x + \epsilon_2 \sin \phi_2 e_y \right],
	\end{aligned}
\end{eqnarray}
with the scalars $a_1$ and $a_2$ being the field amplitudes, the function $g_{1,2}(\phi)$ denoting the slow wave envelopes, and $\epsilon_{1,2}$ being $+1$ or $-1$ corresponding to left or right rotation of the wave, respectively. In the following we use the normalized value $\xi_{1,2}=-ea_{1,2}/m$ for the field strength while $e_x = (0,1,0,0)$ and $e_y = (0,0,1,0)$ are unit vectors. We choose the optical frequency $\omega = 1.55eV$ for the lasers in this work. Furthermore, the classical trajectory is related to the momentum according to
\begin{equation}
	\label{eq:clas.traj}
	\textbf{x}(\tau)= \int d \tau \frac{\textbf{P}(\tau)}{m}.
\end{equation}

\subsection{Analytical solutions}
	\label{sec:traj.ana}
	
The exact solution for \eqref{eq:equ.of.mov} in the background field of \eqref{eq:vec.pot.def} is not feasible. In order to solve it approximately, we set the envelope functions $g_{1,2}(\phi)$ to unity and also make the following assumptions. Firstly, the phases $\phi_1$ and $\phi_2$ in \eqref{eq:vec.pot.def} can be written as 
\begin{eqnarray}
	\label{eq:phase.def}
	\begin{aligned}
    	&\phi_1 = k_1\cdot x(\tau) = \frac{k_1\cdot \bar{P}}{m}\tau + \delta \phi_1(\tau),\\
    	&\phi_2 = k_2\cdot x(\tau) = \frac{k_2\cdot \bar{P}}{m}\tau + \delta \phi_2(\tau),
	\end{aligned}
\end{eqnarray}
with
\begin{eqnarray}
	\label{eq:dphase.def}
	\begin{aligned}
	    \delta\phi_1 = \int\frac{k_1\cdot\delta P(\tau)}{m}d\tau \,,
    	\delta\phi_2 = \int\frac{k_2\cdot\delta P(\tau)}{m}d\tau \,,
	\end{aligned}
\end{eqnarray}
being the higher order corrections. Here, the wave vectors read $k_1 = (\omega,0,0,\omega)$, $k_2 = (\omega,0,0,-\omega)$ and $\delta P(\tau)=P(\tau)-\bar{P}$ with the bar symbol indicating the time averaged quantity in the lasers. 

Another assumption in the derivation allows us to calculate the following integrals
\begin{eqnarray}
	\label{eq:int.phase}
	\begin{aligned}
    	&\int d\tau \sin(\phi_1) \approx -\frac{m}{k_1\cdot \bar{P}} \cos(\phi_1) \,,\\
    	&\int d\tau \sin(\phi_2) \approx -\frac{m}{k_2\cdot \bar{P}} \cos(\phi_2) \,,\\
    	&\int d\tau \sin(\phi_1 - \phi_2) \approx -\frac{m}{(k_1-k_2)\cdot \bar{P}} \cos(\phi_1-\phi_2) \,, \\
    	&\int d\tau \sin(\phi_1 + \phi_2) \approx -\frac{m}{(k_1+k_2)\cdot \bar{P}} \cos(\phi_1+\phi_2) \,,
	\end{aligned}
\end{eqnarray}
as well as the similar integral of $\cos \rightarrow \sin$. By employing these assumptions, the particle's classical equation of motion Eq.~\eqref{eq:equ.of.mov} can be integrated, yielding the classical trajectory.

Since the vector potential in Eq.~\eqref{eq:vec.pot.def} is independent of the transverse coordinates $x$ and $y$, the canonical momenta in these directions are conserved leading to
\begin{equation}
	P_{\bot}(\tau) = p_{\bot}-eA(\tau).
\end{equation}
Without loss of generality, we choose the initial transverse momentum $p_{\bot}$ to be on the $x$-axis. Substituting the explicit vector potential, one arrives at
\begin{eqnarray}
	\label{eq:momen.Px}
	&P_x(\tau)= p_x + m \xi_1 \cos \phi_1 +m \xi_2 \cos \phi_2 \,,\\	
	\label{eq:momen.Py}
	&P_y(\tau)= m \xi_1 \epsilon_1\sin \phi_1 + m \xi_2 \epsilon_2 \sin \phi_2 \,.
\end{eqnarray}
Employing these transverse momenta and the magnetic field components of the laser field 
\begin{eqnarray}
	\label{eq:vec.mag.B}
  	&\textbf{B}_1 = -\omega a_1 (-\epsilon_1 \cos \phi_1 \hat{\textbf{x}} - \sin \phi_1 \hat{\textbf{y}}) \,,\\
    &\textbf{B}_2 =  \omega a_2 (-\epsilon_2 \cos \phi_2 \hat{\textbf{x}} - \sin \phi_2 \hat{\textbf{y}}) \,,
\end{eqnarray}
the equation obeyed by the momentum along the z-direction is
\begin{eqnarray}
	\label{eq:momen.dPz}
	\begin{aligned}
    	\frac{dP_z}{d\tau} =& -p_x\omega \left[\xi_1 \sin \phi_1 - \xi_2 \sin \phi_2 \right] \\
    	                           & -m\xi_1 \xi_2 \omega (1+\epsilon_1\epsilon_2) \sin(\phi_1-\phi_2) \,.
	\end{aligned}
\end{eqnarray}
After integrating over $\tau$, we arrived at
\begin{eqnarray}
	\label{eq:momen.Pz}
	\begin{aligned}
	    P_z = \bar{P}_z &+ p_x\omega \left[ \frac{m\xi_1}{k_1\cdot \bar{P}} \cos \phi_1 -\frac{m\xi_2}{k_2\cdot \bar{P}} \cos \phi_2 \right] \\
	                    &+ \frac{m^2\omega\xi_1\xi_2 \epsilon_-}{(k_1 - k_2)\cdot\bar{P}} \cos(\phi_1-\phi_2)
	\end{aligned}
\end{eqnarray}
with $\epsilon_-=1+\epsilon_1 \epsilon_2$. After having derived the expressions for the momenta in all directions, the energy of the electron in the laser fields can be obtained based on the energy-momentum relation 
\begin{equation}
	\label{eq:en}
    \varepsilon = (m^2 + P_x^2 + P_y^2 + P_z^2)^{1/2} \,,
\end{equation}
with $m_*=m\sqrt{1+\xi_1^2+\xi_2^2}$ being the effective mass of the electron in the laser fields. By defining the time averaged energy as $\bar{\varepsilon}=\sqrt{m_*^2 + p_x^2 + \bar{P}_z^2}$, the energy can be rewritten as 
\begin{equation}
	\label{eq:en.final}
    \varepsilon = \bar{\varepsilon} + \delta \varepsilon + \bar{\varepsilon}{\mathcal{O}}\left(\frac{\delta P_z}{\bar{\varepsilon}} \right)^2
                                                         + \bar{\varepsilon}{\mathcal{O}}\left(\frac{\delta \varepsilon}{\bar{\varepsilon}} \right)^2 \,,
\end{equation} 
where we applied the Taylor expansion. The oscillating term in energy looks like
\begin{equation}
	\label{eq:delta.en}
	\begin{aligned}
    	\delta \varepsilon =& p_x\omega \left(\frac{m\xi_1}{k_1\cdot \bar{P}} \cos \phi_1 + \frac{m\xi_2}{k_2\cdot \bar{P}}\cos\phi_2\right) \\
                            & + \frac{m^2 \omega \xi_1\xi_2 \epsilon_+}{(k_1 + k_2)\cdot\bar{P}} \cos(\phi_1 + \phi_2) \,,
	\end{aligned}
\end{equation}  
with $\epsilon_+=1-\epsilon_1\epsilon_2$. In order for the expansion in Eq.~\eqref{eq:en.final} to be valid, we need ${\mathcal{O}}\left(\frac{\delta P_z}{\bar{\varepsilon}} \right) \ll 1$ and ${\mathcal{O}}\left(\frac{\delta \varepsilon}{\bar{\varepsilon}} \right) \ll 1$, which gives us one condition for our derivation
\begin{equation}
	\label{eq:condition1}
	\begin{aligned}
    	& \frac{m\xi_1 p_x\omega}{k_1\cdot \bar{P} \bar{\varepsilon}} \ll 1 \,, 
    	\frac{m^2\xi_1\xi_2 \omega}{(k_1 + k_2)\cdot\bar{P}\bar{\varepsilon}} \ll 1 \,, \\
    	& \frac{m\xi_2 p_x\omega}{k_2\cdot \bar{P} \bar{\varepsilon}} \ll 1 \,, 
    	\frac{m^2\xi_1\xi_2 \omega}{(k_1 - k_2)\cdot\bar{P}\bar{\varepsilon}} \ll 1 \,. \\
   	\end{aligned}
\end{equation}

\begin{figure*}
	\begin{center}
	\includegraphics[width=0.95\textwidth]{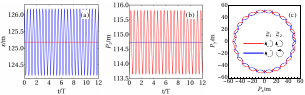}
	\caption{This figure depicts the four-momentum of the electron in the CPW laser fields with $\xi_1=50$ and $\xi_2 = 2.5$. The average energy is chosen to be $\bar{\varepsilon}=125m$ and the initial transverse momentum $p_x = 0$. The electron is on average propagating along with $\xi_1$. In panel (a) and (b), the energy $\varepsilon$ and momentum $P_z$ are shown as a function of the interaction time. In panel (c), the motion of the electron in the $P_x-P_y$ plane is portrayed. The blue curves denote the counter-rotating case ($\epsilon_1\epsilon_2=-1$) while the red for the co-rotating case ($\epsilon_1\epsilon_2=1$). $T=2\pi/\omega$ is the laser period in the laboratory frame.}
	\label{fig:co.vs.counter.all}
	\end{center}
\end{figure*}

So far, we have not given the explicit form for the phases $\phi_1$ and $\phi_2$. But by substituting energy and momentum into Eq.~\eqref{eq:dphase.def}, we can write the higher order terms of the phases as
\begin{eqnarray}
	\label{eq:dphase.exp1}
	\begin{aligned}
	    \delta\phi_1 = \Phi_1 + C_1\sin(\phi_2) &- C_{1-2}\sin(\phi_1-\phi_2) \\
	                                            &+ C_{1+2}\sin(\phi_1+\phi_2) \,,
	\end{aligned}
\end{eqnarray}
\begin{eqnarray}
	\label{eq:dphase.exp2}
	\begin{aligned}
	    \delta\phi_2 = \Phi_2 + C_2\sin(\phi_1) &+ C_{1-2}\sin(\phi_1-\phi_2) \\ 
	                                            &+ C_{1+2}\sin(\phi_1+\phi_2) \,,
	\end{aligned}
\end{eqnarray}
where the coefficients are
\begin{equation}
	\label{eq:dphase.coef}
	\begin{aligned}
		C_1 &= \frac{2p_x m\xi_2\omega^2}{(k_2\cdot\bar{P})^2} \,, 
		C_{1-2} = \frac{\epsilon_- m^2\xi_1\xi_2\omega^2}{[(k_1-k_2)\cdot\bar{P}]^2} \,, \\
		C_2 &= \frac{2p_x m\xi_1\omega^2}{(k_1\cdot\bar{P})^2} \,, 
		C_{1+2} = \frac{\epsilon_+ m^2\xi_1\xi_2\omega^2}{[(k_1+k_2)\cdot\bar{P}]^2} \,.
   	\end{aligned}
\end{equation}
Now, with Eqs.~\eqref{eq:phase.def} and (\ref{eq:dphase.exp1},\ref{eq:dphase.exp2}), we got an implicit system for the solution of the phases. Without loss of generality, we choose here the electron to co-propagate along $\xi_1$, which results in only $C_2$ being not negligible. In order to make sure that the contributions of $C_1$,  $C_{1+2}$, and $C_{1-2}$ to the momentum are of second order and the key assumption in Eq.~\eqref{eq:int.phase} indeed holds, we follow a similar procedure to the one shown in Ref.~\cite{Lv_2021b} and obtain the condition for our solution, which can be written as
\begin{eqnarray}
	\label{eq:cond2}
	\begin{aligned}
    	&\frac{2p_x m\xi_1 \omega^2}{(k_2 \cdot \bar{P})^2} \ll 1 \,,
    	\frac{2p_x m\xi_1 \omega^2}{(k_1 \cdot \bar{P}) (k_2 \cdot \bar{P})} \ll 1 \,, \\
    	&\frac{m^2 \xi_1 \xi_2 \omega}{(k_1 \cdot \bar{P}) \bar{P}_z} \ll 1 \,,
    	\frac{m^2 \xi_1 \xi_2}{\bar{P}_z^2} \ll 1 \,.
	\end{aligned}
\end{eqnarray}
Thus, combining these conditions with the condition in Eq.~\eqref{eq:condition1} yields the closing validity criteria for the solution
\begin{eqnarray}
	\label{eq:cirterion1F}
 	\frac{m^2\xi_2\xi_1}{\varepsilon^2} \ll 1 \,, \\
	\label{eq:cirterion2F}
	\frac{2 p_x m \xi_1}{m_*^2} \ll 1 \,, \\
	\label{eq:cirterion3F}
 	\frac{p_x m \xi_2}{2\varepsilon^2} \ll 1 \,.
\end{eqnarray}

Finally, the classical four-momentum can be written in a covariant form as 
\begin{widetext}
\begin{eqnarray}
	\label{eq:four.momen.final}
	\begin{aligned}
	P^{\mu} &= \bar{P}^{\mu}-e\left[A_1^{\mu}(\phi_1) + A_2^{\mu}(\phi_1)\right] 
	         + k_1^{\mu}\left[\frac{ep \cdot A_1(\phi_1)}{k_1 \cdot \bar{P}} + \frac{\epsilon_+ e^2 A_1^{\mu}(\phi_1) \cdot A_2^{\mu}(\phi_2)}{2(k_1+k_2) \cdot \bar{P}} 
	                                                                         - \frac{\epsilon_- e^2 A_1^{\mu}(\phi_1) \cdot A_2^{\mu}(\phi_2)}{2(k_1-k_2) \cdot \bar{P}}\right] \\ 
	        &+ k_2^{\mu}\left[\frac{ep \cdot A_2(\phi_2)}{k_2 \cdot \bar{P}} + \frac{\epsilon_+ e^2 A_2^{\mu}(\phi_1) \cdot A_2^{\mu}(\phi_2)}{2(k_1+k_2) \cdot \bar{P}} 
	                                                                         + \frac{\epsilon_- e^2 A_2^{\mu}(\phi_1) \cdot A_2^{\mu}(\phi_2)}{2(k_1-k_2) \cdot \bar{P}}\right] \,,	
	\end{aligned}
\end{eqnarray}
\end{widetext}
with $p_\mu$ being the asymptotic momentum and $\epsilon_-=2,\epsilon_+=0$ for the co-rotating case and $\epsilon_-=0,\epsilon_+=2$ for the counter-rotating case. One can verify that in the case of co-rotating, the formula recovers the results in Ref.\cite{Lv_2021b}. Also, if one of the laser beams vanishes, our result recovers the familiar plane wave solution \cite{Ritus_1985}. As the momentum is available now, we can write down the trajectory of the electron as a function of the proper time based on Eq.~\eqref{eq:clas.traj} like 
\begin{widetext}
\begin{eqnarray}
	\label{eq:clas.traj.final}
	\begin{aligned}
    	t &= \frac{\bar{\varepsilon}}{m}\tau + p_x\omega\left(\frac{m\xi_1}{(k_1\cdot \bar{P})^2}\sin\phi_1 + \frac{m\xi_2}{(k_2\cdot \bar{P})^2}\sin\phi_2 \right) 
    	  + \frac{m^2\omega\xi_1\xi_2 \epsilon_+}{[(k_1 + k_2)\cdot \bar{P}]^2}\sin(\phi_1 + \phi_2) \,, \\
	    x &= \frac{p_x}{m}\tau + \frac{m\xi_1}{k_1 \cdot \bar{P}}\sin\phi_1 + \frac{m\xi_2}{k_2 \cdot \bar{P}}\sin\phi_2 \,, \quad	    
	    y = -\frac{m\xi_1\epsilon_1}{k_1 \cdot \bar{P}}\cos\phi_1 - \frac{m\xi_2\epsilon_2}{k_2 \cdot \bar{P}}\cos\phi_2 \,, \\
	    z &= \frac{\bar{P_z}}{m}\tau + p_x\omega\left(\frac{m\xi_1}{(k_1\cdot \bar{P})^2}\sin\phi_1 - \frac{m\xi_2}{(k_2\cdot \bar{P})^2}\sin\phi_2\right)  
	      + \frac{m^2\omega\xi_1\xi_2 \epsilon_-}{[(k_1 - k_2)\cdot \bar{P}]^2}\sin(\phi_1 - \phi_2) \,.
	\end{aligned}
\end{eqnarray} 
\end{widetext} 

Now, the only relation left out is the one between the asymptotic momentum and the average momentum in the laser fields. This relation depends on the turn-on process of the two laser beams \cite{Lv_2021b}. If the beams are turned on separately, one can write down an analytical expression for the average momentum. In the case of the $\xi_1$ beam being turned on first, we have
\begin{equation}
	\label{eq:barP}
 	\bar{P}^{\mu} =p^{\mu}+  \frac{m^2 \xi_1^2}{2( k_1 \cdot p)} k_1^{\mu}+\frac{m^2 \xi_2^2}{2 [k_2 \cdot \bar{P}^{(1)}]} k_2^{\mu},
\end{equation}
where
\begin{equation}
 	\bar{P}^{(1)}_{\mu} =p_{\mu}+ \frac{m^2 \xi_1^2}{2( k_1 \cdot p) } k_{1, \mu}.
\end{equation}
If the $\xi_2$ beam is turned on first, an analogous derivation leads to
\begin{equation}
 \bar{P}^{\mu} =p^{\mu}+  \frac{m^2 \xi_1^2}{2[k_1 \cdot \bar{P}^{(2)}]} k_1^{\mu}+\frac{m^2 \xi_2^2}{2 (k_2 \cdot p)} k_2^{\mu},
\end{equation}
where
\begin{equation}
 \bar{P}^{(2)}_{\mu} =p_{\mu}+ \frac{m^2 \xi_1^2}{2 (k_2 \cdot p )} k_{2, \mu}.
\end{equation}

\subsection{Numerical results}
	\label{sec:traj.num}

With the analytical expressions for the electron's momenta and coordinates being derived, we study in this section the main properties of the motion, with emphasis on the dissimilarities between the cases of co- and counter-rotating laser fields. From Eq.~\eqref{eq:four.momen.final}, we can see that there are two main differences. The first difference is the rotation direction in the $x-y$ plane as $P_y$ changes its direction between co- and counter-rotating cases. The second one is the crossing term proportional to $\xi_1 \xi_2$. For the co-rotating case, the crossing term emerges in the momentum $P_z$ while in the counter-rotating case it  appears in the energy $\varepsilon$.

\begin{figure*}
	\begin{center}
	\includegraphics[width=0.95\textwidth]{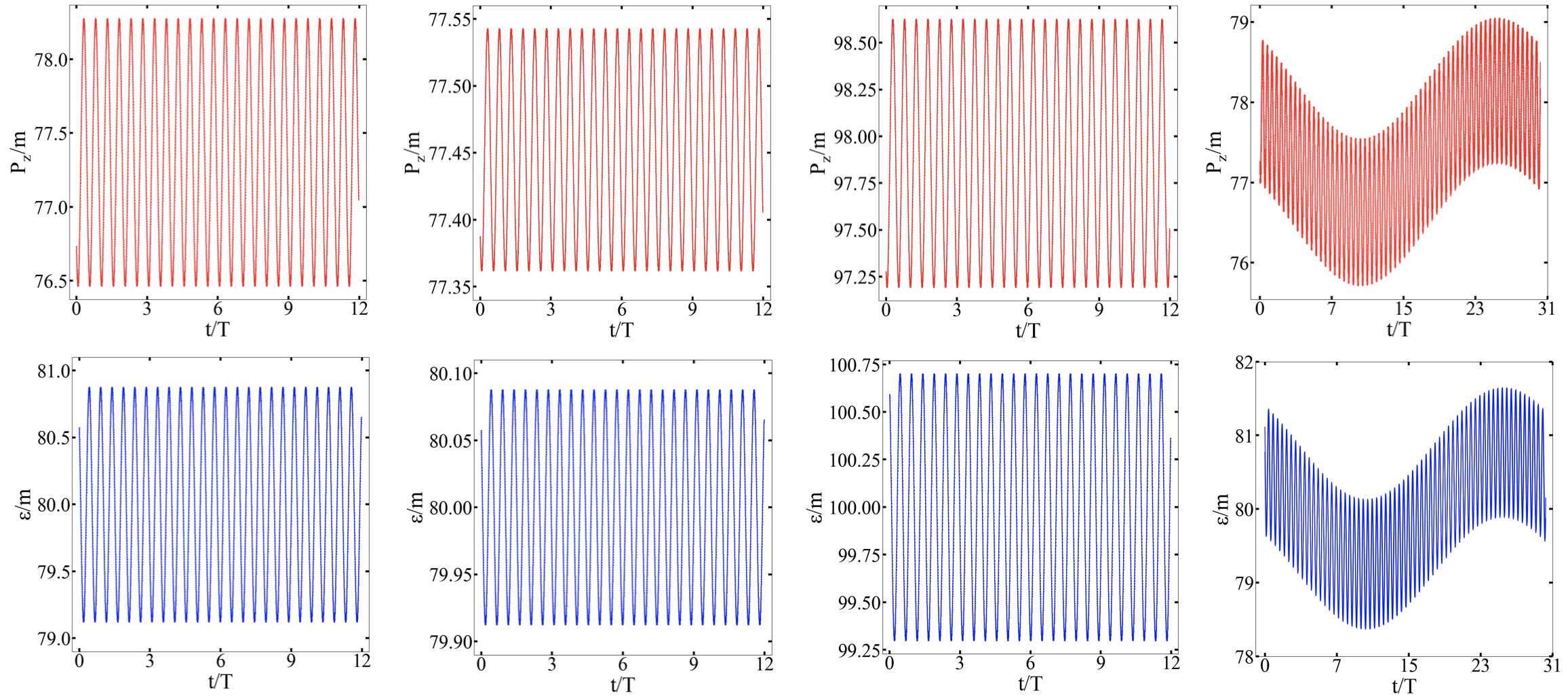}
	\caption{The momentum $P_z$ for the co-rotating case and the energy $\varepsilon$ for the counter-rotating case are displayed as a function of the interaction time. The first row is for co-rotating case (red) while the second row is for counter-rotating case (blue). The solid lines are the analytical results based on Eq.~\eqref{eq:four.momen.final} and the dots are the numerical solution of Eq.~\eqref{eq:equ.of.mov}. In all the panels, we choose $\xi_1=20$. The strength for the second laser is chosen to be $\xi_2=3.5$ except for the second column where $\xi_2=0.35$. The average energy is chosen to be $\bar{\varepsilon}=80m$ for the first, second, and last column while for the third column we choose $\bar{\varepsilon}=100m$. The transverse momentum $p_x$ is $0.1m$ for the last column otherwise zero.} 
	\label{fig:co.vs.counter.PzEn}
	\end{center}
\end{figure*}

We inspect the electron's four-momentum for both co-rotating and counter-rotating cases in Fig.~\ref{fig:co.vs.counter.all}, where we cover different behaviors between the two cases. The results shown in the figure are based on Eq.~\eqref{eq:four.momen.final} but proved by the fully numerical solutions of Eq.~\eqref{eq:equ.of.mov}. The average energy is $\bar{\varepsilon}=125m$ , corresponding to $\omega_2/\omega_1=22.95$. Here, $\omega_{1,2} \equiv k_{1,2}\cdot\bar{P}/\bar{\varepsilon}$ is the laser frequency in the electron's rest frame. The plots represent a time interval of one cycle of the $\xi_1$ beam and 23 cycles of the $\xi_2$ beam as the electron is on average propagating along with $\xi_1$.

By choosing a vanishing transverse momentum, we can see, from panel (a), that the energy is a constant for the co-rotating case while oscillating around the same constant for the counter-rotating one. The amplitute of the oscillation is proportional to $\xi_1\xi_2/\bar{\varepsilon}$ and the frequncy of the oscillation is approximately $\omega_2$ as $\omega_2/\omega_1 \gg 1$. On the other hand, the behavior for the momentum along the $z$-direction in panel (b) is exactly swapped. $P_z$ for the co-rorating case is oscillating around the constant $P_z$ of the counter-rotating case with an amplitude proportional to $\xi_1\xi_2/\bar{P}_z$. The frequency is the same as the one for the oscillations in the energy [panel (a)] since from Eq.~\eqref{eq:four.momen.final} we can see the crossing terms are both dependent on $\phi_1$ and $\phi_2$ with the way bigger $\phi_2$ dominating over the smaller $\phi_1$.

In panel (c) of Fig.~\ref{fig:co.vs.counter.all}, the momentum in the tranverse plane is displayed. Eq.~\eqref{eq:four.momen.final} shows us that oscillations in $P_x$ and $P_y$ consist of two parts. One of them is related to the $\xi_1$-laser. Since the electron is propagating along with $\xi_1$, this oscillation gives the large enclosing circle, which is of size $m \xi_1$. The small amplitude and high frequency oscillations caused by the $\xi_2$-laser are then giving the small kinks and arcs along the big $\xi_1$ circle. The number of these arcs is related to $\omega_2/\omega_1$ as it depends on how many $\xi_2$ oscillations the electron is cycling through during one $\xi_1$ oscillation and therefore depends linearly on the energy of the electron. The radius of the arc, on the other hand, depends on $m \xi_2$.

To show the accuracy of our analyical solutions, we compared the analytical result with the numerical one in Fig.~\ref{fig:co.vs.counter.PzEn}. It has been shown in Eq.~\eqref{eq:four.momen.final} as well as in Fig.~\ref{fig:co.vs.counter.all} that the higher order oscillation, namely the crossing term, only appears in $\varepsilon$ for the counter-rotating case and in $P_z$ for the co-rotating case. Therefore, we plot the energy $\varepsilon$ for the counter-rotating case and the $P_z$-momentum for the co-rotating case in Fig.~\ref{fig:co.vs.counter.PzEn}. 

Comparing the first column with the second one, we can see that the oscillation amplitude decreases by about one order of magnitude while the average $P_z$ (co-rotating) or $\varepsilon$ (counter-rotating) stays the same. This is because $\xi_2$ becomes one order of magnitude smaller and the crossing term is proportional to it, see in Eq.~\eqref{eq:four.momen.final}. By increasing the energy $\varepsilon$ from $80m$ (first column) to $100m$ (third column), the oscillation amplitude decreases again because the prefactor of the crossing term is proportional to $1/\varepsilon$ in the counter-rotating case and $1/P_z$ in the co-rotating case. In the last column, the transverse momentum $p_x$ has been introduced and we chose $p_x=0.1m$. We can see now two kinds of oscillations in both $\varepsilon$ and $P_z$. This is because if $p_x \neq 0$ the oscillation related solely to $\xi_1$ or $\xi_2$ in $\varepsilon$ and $P_z$ appears additionally to the crossing term. The fast oscillation is now the result of two individual oscillations, which relate to $\xi_2$ and the crossing term of $\xi_1 \xi_2$, respectively.

In order to quantitatively characterize the disagreement between the numerical and analytical results, we have introduced the relative deviation of the analytical prediction (subscript $a$) of a quantity $X$ with respect to the numerically calculated value (subscript $n$) as follows
\begin{equation}
	\label{eq:Delta_def}
	\Delta_X \equiv \frac{1}{2T}\int_{-T}^{T} dt \left| \frac{X_{a}-X_{n}}{X_{n}} \right|,
\end{equation}
with $X$ being either the longitudinal momentum $P_z$ for the co-rotating case or the energy $\varepsilon$ for the counter-rotating case. The integration time is taken to infinity, i.e. $T \rightarrow \infty$. It should be noticed that the phases $\phi_1,\phi_2$ contain arbitrary constants $\Phi_1,\Phi_2$. In order to compare the analytical and numerical quantities these constants should be specified. We write down $\Phi_1$ and $\Phi_2$ as $\Phi_1=k_1 \cdot x_0$ and $\Phi_2=k_2 \cdot x_0$ where $x_0$ is the temporal and spatial location of the particle at the moment when the turn-on process in the numerical simulation has finished. In this case we can have the same initial phase for both analytical and numerical solutions. The relative difference for the four cases shown in Fig.~\ref{fig:co.vs.counter.PzEn} has been calculated together with the conditions (\ref{eq:cirterion1F} $\sim$ \ref{eq:cirterion3F}) of the analytical solutions in Tab.~\ref{tab:table}.   
  
\begin{table*}
	\begin{center}
    \begin{tabular}{|c|c|c|c|}
      \hline
      \multirow{2}{*}{case} & criteria & \multirow{2}{*}{$\Delta_{P_z}$} & \multirow{2}{*}{$\Delta_{\varepsilon}$} \\
       & $\bigl[$Eq.~\eqref{eq:cirterion1F},Eq.~\eqref{eq:cirterion2F},Eq.~\eqref{eq:cirterion3F}$\bigr]$ &  & \\
      \hline
       1st column & [$0.011,0.0,0.0$] & $1.05 \times 10^{-4}$ & $7.84 \times 10^{-5} $ \\
      \hline
       2nd column & [$0.001,0.0,0.0$] & $2.87 \times 10^{-6}$ & $7.68 \times 10^{-7} $ \\
      \hline
       3rd column & [$0.007,0.0,0.0$] & $4.68 \times 10^{-5}$ & $3.58 \times 10^{-5} $ \\
      \hline
       4th column & [$0.011,0.010,0.00003$] & $2.10 \times 10^{-4}$ & $1.50 \times 10^{-4} $ \\
      \hline
    \end{tabular}
    \caption{The relative difference in the energy $\Delta_{\varepsilon}$ for the counter-rotating case and in the longitudinal momentum $\Delta_{P_z}$ for the co-rotating case with same parameters as in Fig.~\ref{fig:co.vs.counter.PzEn}.}
    \label{tab:table}
  \end{center}
\end{table*}

From Tab.~\ref{tab:table}, we can see that the relative difference between the numerical solution and the analytical solution is quite small within the parameter regime we have considered and increases if the criteria are not well fulfilled anymore, as for example when the transverse momentum is introduced. This small relative difference with respect to the exact numerical solution gives us the justification to employ the analytical solution to study the radiation of an ultrarelativistic electron in a CPW setup within certain parameter regimes.

\section{quantum radiation}
\label{sec:radi}

\subsection{Radiation formulas}
	\label{sec:radi.ana}

Applying the classical trajectory, the emission can be calculated according to the Baier-Katkov method\cite{Katkov_1968,Baier_b_1994, Landau_4}. For the sake of simplicity, we start with a spinless particle. An analogous derivation for the spinor case is given later. The Baier-Katkov expression for the emitted intensity $dI$ reads
\begin{equation}
	\label{eq:inten.BK0}
	dI = \frac{\alpha \varepsilon}{(2 \pi)^2 \varepsilon'T_0} |\mathcal{T}_{\mu}|^2 d^3k'
\end{equation}
where $\alpha$ is the fine structure constant, $T_0$ is the interaction time, and $\varepsilon' = \varepsilon - \omega'$ is the energy of the electron after the emission of a photon. The transition amplitude 
\begin{equation}
	\label{eq:T.def.t}
	\mathcal{T}_{\mu}(k') = \int_{-\infty}^{\infty}{dt}v_{\mu}(t) e^{i \psi}, \quad \quad \psi \equiv \frac{\varepsilon}{\varepsilon'}k' \cdot x(t),
\end{equation}
with $v_{\mu}=dx_\mu / dt$. $k'_{\mu}$ is the emitted photon four-momentum, characterized by its energy $\omega'$ and the emission direction $\textbf{n} = \left( \cos \varphi \sin \theta, \sin \varphi \sin \theta, \cos \theta \right)$ as
\begin{equation}
	\label{eq:kp.def}
	k'_{\mu} = \omega' \left(1, \textbf{n} \right) \,.
\end{equation}

Within the realm of this theory, the oscillation of $\delta \varepsilon$ is assumed to be small compared to $\varepsilon$, which holds in our case as shown in the previous section. Accordingly, the factor appearing in the phase may be approximated as $\frac{\varepsilon}{\varepsilon'} \approx \frac{\bar{\varepsilon}}{\bar{\varepsilon}'}\left[ 1+\frac{(\delta \varepsilon)^2}{\bar{\varepsilon} \bar{\varepsilon}'} \right]$. In the following derivation the second order correction is neglected. Moreover, for reasons of simplicity, the average energy $\bar{\varepsilon}$ is replaced from now on by $\varepsilon$. Since the trajectory presented in Sec.~\ref{sec:traj} is given in terms of the proper time $\tau$, we change the integration variable in Eq.~\eqref{eq:T.def.t}, leading to
\begin{equation}
	\label{eq:T.def.tau}
	\mathcal{T}_{\mu}(k') = \int_{-\infty}^{\infty}{d \tau} \frac{P_{\mu}(\tau)}{m} e^{i \psi}.
\end{equation}
where the relation between $P_{\mu}$ and $dx_{\mu}/d \tau$ was used. Substituting the trajectory \eqref{eq:clas.traj.final} and the emitted wavevector \eqref{eq:kp.def} into expression \eqref{eq:T.def.tau}, the phase reads
\begin{eqnarray}
	\label{eq:T.phase1}
	\begin{aligned}
		\psi =& \psi_{l} \tau - z_{11}\sin\phi_1 -z_{12}\cos\phi_1 \\
		      &- z_{21} \sin\phi_2 - z_{22} \cos\phi_2 \\
		      &- z_3^{\pm} \sin(\phi_1 \pm \phi_2) \,,
	\end{aligned}
\end{eqnarray}
where $+$ and $-$ correspond to the counter-rotating and co-rotating case respectively and the following quantities were introduced
\begin{eqnarray}
	\label{eq:zs}
	\begin{aligned}
	    z_{11} &= -\frac{u m \xi_1}{\omega_1}\left[-n_x + \frac{p_x\omega}{\varepsilon\omega_1}(1-n_z)\right]   \,, 
    	z_{12} = -\frac{u m \xi_1 \epsilon_1}{\omega_1} n_y   \,, \\
    	z_3^+ &= -\frac{u \omega m^2 \xi_1\xi_2\epsilon_+}{\varepsilon (\omega_1 + \omega_2)^2}  \,,
    	z_3^- = \frac{u \omega m^2 \xi_1\xi_2\epsilon_- n_z}{\varepsilon (\omega_1 - \omega_2)^2}   \,, \\
    	z_{21} &= -\frac{u m\xi_2}{\omega_2}\left[-n_x + \frac{p_x \omega}{\varepsilon\omega_2}(1+n_z)\right]   \,,
    	z_{22} = -\frac{u m\xi_2\epsilon_2}{\omega_2} n_y \,,
	\end{aligned}
\end{eqnarray}
with $u \equiv \omega'/(\varepsilon - \omega')$. By utilizing the following definitions 
\begin{eqnarray}
	\label{eq:z12}
	z_1 &=& \sqrt{z_{11}^2 + z_{12}^2} \,,
	z_2 = \sqrt{z_{21}^2 + z_{22}^2} \,, \\
	\label{eq:varphi12}
 	\varphi_1 &=& \tan^{-1} \left( \frac{z_{12}}{z_{11}}\right) \,, 
  	\varphi_2 =\tan^{-1} \left( \frac{z_{22}}{z_{21}}\right) \,,
\end{eqnarray}
the phase can be simplified even more to
\begin{equation}
	\label{eq:T.phase2}
    \psi = \psi_l \tau - z_1\sin(\phi_1 - \varphi_1) - z_2\sin(\phi_1 - \varphi_1) - z_3^{\pm} \sin(\phi_1 \pm \phi_2) \,.
\end{equation}
The linear term in the phase has the coefficient
\begin{equation}
	\label{eq:psi_np}
    \psi_l = \frac{\varepsilon^2 u}{m}\left(1 - \frac{v_x}{\varepsilon}n_x - \frac{v_z}{\varepsilon}n_z \right)  \,.
\end{equation}
with $v_x=p_x/\varepsilon$ and $v_z=\bar{P}_z/\varepsilon$ as the average velocities along $x$ and $z$ directions. Substituting Eq.~\eqref{eq:T.phase2} as well as Eq.~\eqref{eq:four.momen.final} into Eq.~\eqref{eq:T.def.tau} and using the Jacobi-Anger expansion for the Bessel function, we can obtain the transition amplitudes after some tedious but straightforward derivations
\begin{equation}
	\mathcal{T}_{\mu} = 2\pi \sum_{s_1,s_2,s_3} \mathcal{M}_{\mu}(s_1,s_2,s_3) \delta(\Omega_{s_1,s_2,s_3}),
\end{equation}
where the $\delta$ function argument is given by
\begin{equation}
	\label{eq:Om_rl}
	\Omega_{s_1,s_2,s_3} \equiv \psi_l - \frac{\varepsilon}{m} \left[ (s_1+s_3) \omega_1 + (s_2 \pm s_3) \omega_2 \right] \,,
\end{equation}
with $+$ and $-$ in the last term representing the counter-rotating and the co-rotating case, respectively. One may notice that different combinations of the indices $s_1,s_2,s_3$ may yield the same $\delta$ function argument. As a result, when squaring $\mathcal{T}$ interference terms will arise but depend on the quantity $\omega_2/\omega_1$. If this ratio is an integer, the motion is periodic with the frequence $2 \pi/ \omega_1$. Otherwise, the motion is non-periodic. Because the periodic motion not easily fulfilled in reality, we will, therefore, focus on the non-periodic motion in the study below.  

By defining $s_R \equiv s_1+s_3$ and $s_L \equiv s_2 \pm s_3$, one may write
\begin{equation}
	\label{eq:T.final}
	\mathcal{T}_{\mu} = 2 \pi \sum_{s_L, s_R} \mathcal{M}_{\mu} (s_L,s_R) \delta(\Omega_{s_L,s_R}) \,.
\end{equation}
with $\Omega_{s_R,s_L} \equiv \psi_l - \varepsilon/m \left( s_R \omega_1 + s_L \omega_2 \right)$. The matrix elements take the form
\begin{widetext}
\begin{eqnarray}
	\label{eq:M0.final}
	\mathcal{M}_0 &=& \sum_{s_3} \left[\left(\frac{\varepsilon}{m} B_0(\textbf{3}) + \frac{\epsilon_+ \omega m \xi_1 \xi_2}{\varepsilon (\omega_1+\omega_2)} B_1(\textbf{3})\right) B_0(\textbf{1}) B_0(\textbf{2}) + p_x \omega \left(\frac{\xi_1}{\varepsilon\omega_1} B_1(\textbf{1}) B_0(\textbf{2}) + \frac{\xi_2}{\varepsilon\omega_2} B_0(\textbf{1}) B_1(\textbf{2})\right)B_0(\textbf{3})  \right]\,,\\	
	\label{eq:M1.final}
	\mathcal{M}_1 &=& \sum_{s_3}\left[ \frac{p_x}{m} B_0(\textbf{1}) B_0(\textbf{2}) B_0(\textbf{3}) + \biggl(\xi_1 B_0(\textbf{2})B_1(\textbf{1}) + \xi_2 B_0(\textbf{1})B_1(\textbf{2}) B_0(\textbf{3}) \biggr) \right]\,, \\
	\label{eq:M2.final}
	\mathcal{M}_2 &=& \sum_{s_3} \biggr[ \xi_1\mu_1 B_0(\textbf{2})B_2(\textbf{1}) + \xi_2 \mu_2 B_0(\textbf{1})B_2(\textbf{2}) \biggl] B_0(\textbf{3}) \,,\\
	\label{eq:M3.final}
	\mathcal{M}_3 &=& \sum_{s_3} \left[\left(\frac{\bar{P}_z}{m} B_0(\textbf{3}) - \frac{\epsilon_- \omega m \xi_1 \xi_2 }{\varepsilon (\omega_1-\omega_2)} B_1(\textbf{3}) \right) B_0(\textbf{1}) B_0(\textbf{2}) + p_x \omega  \left(\frac{\xi_1}{\varepsilon \omega_1} B_1(\textbf{1}) B_0(\textbf{2}) -\frac{\xi_2}{\varepsilon\omega_2} B_0(\textbf{1}) B_1(\textbf{2})B_0(\textbf{3})  \right) \right].
\end{eqnarray}
\end{widetext}
Here, we have $\textbf{1} \equiv (s_1,z_1,\varphi_1)$, $\textbf{2} \equiv (s_2,z_2,\varphi_2)$, and $\textbf{3} \equiv (s_3,z_3,0)$. In the derivation, we considered the identities used in Ref. \cite{Ritus_1985}
\begin{equation}
	(1,\cos \phi, \sin \phi) e^{-z \sin \left( \phi- \varphi \right)} = \sum_s (B_0,B_1,B_2) e^{-is \phi}.
\end{equation}
The functions $B_0,B_1,B_2$ are related to the Bessel function and its first derivative $J_s(z),J'_s(z)$ through
\begin{eqnarray}
	B_0(s,z,\varphi) &=& J_s(z)e^{is \varphi} \,, \\
	B_1(s,z,\varphi) &=& \left[ \frac{s}{z}J_s(z) \cos \varphi - i J'_s(z) \sin \varphi \right] e^{i s \varphi} \,, \\
	B_2(s,z,\varphi) &= &\left[ \frac{s}{z}J_s(z) \sin \varphi + i J'_s(z) \cos \varphi \right] e^{i s \varphi} \,.
\end{eqnarray}

Finally, the emitted intensity may be obtained by integrating (\ref{eq:inten.BK0}) over the polar angle.
\begin{multline}
	\label{eq:emis1}
	\frac{dI}{d \omega' d \varphi} = \frac{\alpha m }{ 2 \pi \varepsilon'} \int d(\cos \theta) \omega'^2 \sum_{s_L, s_R} \Bigl| \mathcal{M}_{\mu} (s_L,s_R)\Bigr|^2 \delta(\Omega_{s_L,s_R})\,,
\end{multline}
where the identity $\delta^2(\Omega_{s_L,s_R})=\frac{\tau_0}{2 \pi}\delta(\Omega_{s_L,s_R})$ has been used. The proper interaction time is given by $\tau_0 = (m/\varepsilon) T_0$. As squaring $\mathcal{T}$ does not mix terms associated with different $s_L,s_R$ indices, the interference takes place only between terms included within $\mathcal{M}_{\mu}(s_L,s_R)$. The condition imposed by the $\delta $ function, namely $\Omega_{s_L,s_R} = 0$, illustrates the energy conservation in the radiation process and determines the relation between $\cos \theta$ and $\omega',\varphi$
\begin{equation}
	\label{eq:cos1}
	1-\rho - \bar{v}_z \cos \theta = \bar{v}_x \cos \varphi \sqrt{1-\cos^2 \theta}.
\end{equation}
By squaring and solving this equation one obtains two possible angles
\begin{equation}
	\label{eq:cos2}
	\cos \theta_{\pm} = \frac{\bar{v}_z (1-\rho) \pm \bar{v}_x \cos \varphi \sqrt{\Delta}}{\bar{v}_z^2+\bar{v}_x^2 \cos^2 \varphi} \,,
\end{equation}
where the following quantities were introduced
\begin{eqnarray}
	\label{eq:rho_def}
	\Delta & \equiv & \bar{v}_z^2+\bar{v}_x^2 \cos^2 \varphi - \left(1-\rho \right)^2, \\
	\rho & \equiv & \frac{\left(s_R \omega_1 + s_L \omega_2 \right)}{u \varepsilon} .
\end{eqnarray}
Please note $\pm$ just represents different solutions for $\cos \theta$ and does not relate to the co- or counter-rotating cases. Notice that when squaring (\ref{eq:cos1}) a redundant solution may be added, which solves the equation
\begin{equation}
	\label{eq:cos1temp}
	1-\rho - \bar{v}_z \cos \theta =- \bar{v}_x \cos \varphi \sqrt{1-\cos^2 \theta},
\end{equation}
rather than the original one. Thus, the solutions given in (\ref{eq:cos2}) are physical only if when substituted into the right hand side of (\ref{eq:cos1}), a positive result follows. A solution that does not meet this criterion is therefore excluded. Employing the $\delta$ function to perform the integration leads to
\begin{equation}
	\label{eq:dIdphi}
	\frac{dI}{d \omega' d \varphi} = \frac{\alpha \varepsilon \omega'^2}{ 2 \pi \varepsilon' m} \sum_{i=\pm} \sum_{s_L, s_R} \Bigl| \mathcal{M}_{\mu}(s_L,s_R) \Bigr|^2 \Bigl| 
	                                 \frac{d \Omega_{s_L,s_R}}{d(\cos \theta)} \Bigr|^{-1}_{\theta=\theta_i}.
\end{equation}
The reciprocal of the derivative of the $\delta$ function, required for the integration, reads
\begin{equation}
	\label{eq:deriva}
	\Bigl| \frac{d \Omega_{s_L,s_R}}{d(\cos \theta)} \Bigr|^{-1} = \frac{m}{\varepsilon^2 u}\left| \frac{1}{ \bar{v}_x \cos \varphi \cot \theta -\bar{v}_z } \right| \equiv \kappa(\theta)\,.
\end{equation}
Substituting (\ref{eq:deriva}) into (\ref{eq:dIdphi}) yields the final result
\begin{equation}
	\label{eq:dIdphi.final.bos}
	\frac{dI}{d \omega' d \varphi} = \frac{\alpha m \omega'^2}{ 2 \pi \varepsilon \varepsilon'} \sum_{i=\pm} \sum_{s_L, s_R} \Bigl| \mathcal{M}_{\mu}(s_L,s_R) \Bigr|^2 \kappa(\theta_i) \,.
\end{equation}
Here $i=\pm$ represent the two solutions for $\cos \theta$ in Eq.~\eqref{eq:cos2}.

For a spinor particle the initial emission expression (\ref{eq:inten.BK0}) is modified as follows
\begin{equation}
	|\mathcal{T}|^2 \rightarrow |\mathcal{K}|^2 \equiv -\left( \frac{\varepsilon'^2 + \varepsilon^2 }{2 \varepsilon \varepsilon'} \right) |\mathcal{T}_{\mu}|^2 + \frac{\omega'^2}{2 \varepsilon'^2 \varepsilon'^2} |\mathcal{T}_{0}|^2 \,.
\end{equation}
Therefore, the final results for a spin-$\frac{1}{2}$ particle is obtained
\begin{multline}
	\label{eq:dIdphi.final.fer}
	\frac{dI}{d \omega' d \varphi} = \frac{\alpha m \omega'^2}{2 \pi  \varepsilon} \sum_{i=\pm} \sum_{s_L,s_R} \kappa(\theta_i) \times \\ 
	                                 \left[-\left( \frac{\varepsilon'^2 + \varepsilon^2 }{2 \varepsilon \varepsilon'} \right)|\mathcal{M}_{\mu}(s_L,s_R)|^2 + \frac{\omega'^2}
	                                 {2 \varepsilon'^2 \varepsilon'^2} |\mathcal{M}_0(s_L,s_R)|^2 \right]\,.
\end{multline} 

In the derivation above, we have only included the linear dependence on $\tau$ for the phases $\phi_1$ and $\phi_2$ in the classical momentum Eq.~\eqref{eq:four.momen.final} and trajectory Eq.~\eqref{eq:clas.traj.final} and ignored the higher order corrections. From Eqs.~\eqref{eq:dphase.exp1} and \eqref{eq:dphase.exp2}, we can see that the next order corrections are oscillations with the amplitute of $C_1$, $C_2$, $C_{1-2}$, and $C_{1+2}$ in Eq.~\eqref{eq:dphase.coef}. The difference between the co- and counter-rotating cases is $C_{1-2}$ and $C_{1+2}$, which have the same order of magnitude. Therefore, by following the same procedure as in the previous work \cite{Lv_2021b}, we can infer the same validity conditions for the matrix elements in Eqs.~\eqref{eq:M0.final}$\sim$\eqref{eq:M3.final}, see Eq.~(111) in Ref.\cite{Lv_2021b}.

\subsection{Numerical results}
	\label{sec:radi.num}

In the following we present typical spectra for an ultrarelativistic electron in the strong field regime ($\xi_1 \gg 1$) and we will focus on the influence of the sense of rotation of the laser fields on the electron's radiation spectra. We know that the quantum parameter $\chi= e\sqrt{-(F^{\mu \nu} P_{\nu})^2}/m^3$ can totally characterize the radiation property if LCFA is applied, but in a general field configuration where LCFA may not be fully valid $\chi$ can still characterize some aspects of the radiation. From the previous study \cite{Lv_2021a} we know that there are three different regimes where the radiation behaviour changes dramatically. Therefore, we will also investigate the influence of the co-rotating and counter-rotating laser beams on the radiation spectra in these three regimes. In order to distinguish the different regimes, we also defined $\chi_{1,2} \equiv \xi_{1,2} k_{1,2}\cdot\bar{P}/m^2 $ as the quantum parameters regarding the $\xi_1$- and $\xi_2$-laser, respectively. 

From Eqs.~\eqref{eq:M0.final}$\sim$\eqref{eq:M3.final} for the matrix elements, we can see that the differences between the co-rotating and counter-rotating cases appear in four places: 

(i) The crossing term in $\mathcal{M}_{\mu}$, proportional to $\xi_1 \xi_2$, moves from $\mathcal{M}_3$ in the co-rotating case to $\mathcal{M}_0$ in the counter-rotating case. This is because of the crossing term in the classical four-momentum appearing in energy for the co-rotating case while appearing in $P_z$ for the counter-rotating case, see Eq.~\eqref{eq:four.momen.final}; 

(ii) Two terms in $\mathcal{M}_2$ show a different sign with respect to each other in the counter-rotating case. This is related to the oscillations in $P_y$ changing their direction between co- and counter-rotating cases;

(iii) One of the arguments for the Bessel function $z_3^{\pm}$ in Eq.~\eqref{eq:zs} shows a difference between the two cases. The reason is again due to the crossing term in the four-momentum since this term in the co- or counter-rotating case has a different prefactor. 

(iv) For the Bessel function with the argument $z_2$ the order is also changing between the two cases. From Eq.~\eqref{eq:Om_rl}, we can see that $s_R$ and $s_L$ are the same for both cases when the emitted photon $k'_{\mu}$ is fixed. However, $s_L= s_2 \pm s_3$, depends on $s_2$ and $s_3$ differently for co- and counter-rotating cases, which changes the physical meaning of $s_3$. In the co-rotating case $s_3$ represents the process of absorbing a certain amount of photons from one laser and then emitting the same amount to the other laser, while in the counter-rotating case $s_3$ corresponds to either absorbing from or emitting to both lasers the same amount of photons at once, depending on $s_3$ being positive or negative. 

From Eqs.~\eqref{eq:M0.final}$\sim$\eqref{eq:M3.final}, we can see that the terms proportional to the transverse momentum $p_x$ are the same for both co-rotating and counter-rotating cases and will not contribute to dissimilarities between the two cases. Therefore, we choose $p_x=0$ in the following calculations but the conclusion will be the same for $p_x \neq 0$. Moreover, when $p_x=0$ the validity conditions for the matrix elements will be fulfilled automatically \cite{Lv_2021b}. 

\begin{figure}[t]
  \begin{center}
  \includegraphics[width=0.48\textwidth]{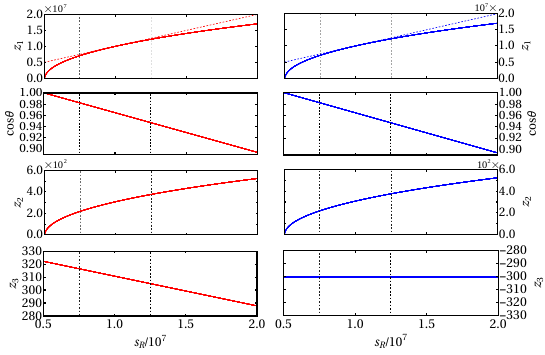}
  \caption{The emission direction $\cos\theta$ and the arguments of the Bessel function $z_1$, $z_2$, and $z_3$ as a function of $s_R$ defined above. The left column (red) is for the co-rotating case while the right one (blue) is for the counter-rotating case. Here, $\xi_1=200$ and $\xi_2=0.35$ and the average energy $\varepsilon=767m$, which corresponds to $\chi_1/chi_2=10$. The emitted photon energy $\omega'$ is chosen to be corresponding to $u=0.02$. The dased lines in the first row correspond to $z_1=s_R$.}
  \label{fig:function.sR}
  \end{center}
\end{figure}

In Regime I ($\chi_1 \gg \chi_2$), the $\xi_1$-laser will dominate the radiation process. The major task in calculating the spectrum is to evaluate the Bessel functions in the matrix elements. We know that $J_n(z)$ will vanish if its order $n$ is larger enough than the argument $z$. Therefore, it is wise to estimate the maximum value of the Bessel function's argument before the calculation of the matrix elements. For this purpose Fig~\ref{fig:function.sR} is illustrating the behaviour of said arguments and $cos(\theta)$ in dependence on $s_R$. It shows that the arguments $z_1$ and $z_2$ and the angle $\theta$ are the same for both case as expected. The only difference lies in $z_3$, which depends on the emission angle in the co-rotating case but is a constant in the counter-rotating case, see also in Eq.~\eqref{eq:zs}. 

Because $\xi_1=200$ and $\varepsilon=767m$ in the calculation, $z_1$ is rather large and increases with the emission angle. However, from the two panels in the first row of Fig.~\ref{fig:function.sR}, we can see that there is a certain region (between the two vertical dashed lines in the plots) when $z_1 \sim s_1$ and the contribution from $J_{s_1}(z_1)$ in the matrix elements is not negligible. Please note that $s_1=s_R-s_3$ and $s_3$ is usually much smaller than $s_R$ as $z_3 \ll z_1$. In this region, $\cos\theta$ is around $0.965$, which coincides with the propagation direction of the electron $P_z/\varepsilon \approx 0.9654$ as expected.

For the third row in Fig.~\ref{fig:function.sR}, we can see that $z_2$ is the same for both co- and counter-rotating cases. However, $s_2$ is different between the two cases as mentioned before. Naturally now the question arises, if this difference in $s_2$ will infer a difference in spectra. Our conjecture is no because the sum over $s_3$ in Eqs.~\eqref{eq:M0.final}$\sim$\eqref{eq:M3.final} will cover the whole region within which $J_{s_2}(z_2)$ is not negligible. The only effect will be the peak of the harmonic in the spectrum maybe shifted when the harmonic structure is obvious. 

From Eq.~\eqref{eq:zs} we can estimate that $z_3^-$ for the co-rotating case and $z_3^+$ for the counter-rotating one have similar amplitudes for the relevant region (see also in the fourth row of Fig.~\ref{fig:function.sR}), even though $z_3^-$ depends on the direction of emission. The different sign between $z_3^-$ and $z_3^+$ will not play a major role when summing over $s_3$, and only a quantitative difference may appear in the spectra.   

\begin{figure}[t]
  \begin{center}
  \includegraphics[width=0.49\textwidth]{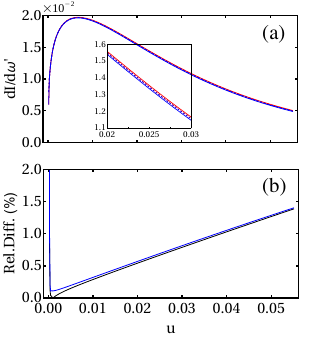}
  \caption{Panel (a): the radiation spectra as a function of $u$ for both co-rotating case (red) and counter-rotating case (blue). The inset is a zoom-in for $0.02<u<0.03$. The solid curves is the results based on Eq.~\eqref{eq:dIdphi.final.fer} and the dashed lines are the prediction of LCFA. Panel (b): the relative difference betweem the accurate results (solid lines) and the LCFA predictions (dashed lines). The parameters for the laser beams and the electron are the same as in Fig.~\ref{fig:function.sR}.}
  \label{fig:spectra200-0.35}
  \end{center}
\end{figure}

To test our above conjectures, the spectra for both co- and counter-rotating cases are displayed in Fig.~\ref{fig:spectra200-0.35}. From panel (a), we can see that the spectra for both cases have exactly the same shape and the difference is almost invisible. Only if zoomed-in, we can see that the co-rotating case gives slightly larger values than the counter-rotating one. More interestingly, the LCFA formula \cite{Ritus_1985} predicts almost the same spectrum for both cases, see in the zoomed-in scale. We have also calculated the relative difference between the accurate spectrum and the LCFA one in panel (b), illustrating that the two cases have a similar relative difference for the main part of the spectrum where the value is around $1\%$. The relative difference is only very large at the very beginning, which is because the harmonic structure at low energies cannot be reproduced by the LCFA.

\begin{figure}[t]
  \begin{center}
  \includegraphics[width=0.49\textwidth]{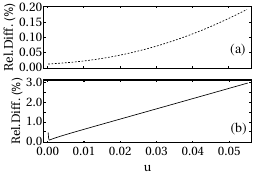}
  \caption{The relative difference between the LCFA results for the two cases in panel (a) and the relative difference between the accurate results in panel (b) for the spectra shown in Fig.~\ref{fig:spectra200-0.35}.}
  \label{fig:rela.diffs}
  \end{center}
\end{figure}

Instead of the relative difference between the accurate spectrum and the LCFA prediction shown in panel (b) of Fig.~\ref{fig:spectra200-0.35}, Fig.~\ref{fig:rela.diffs} depicts the relative difference between the two accurate spectra in Fig.~\ref{fig:spectra200-0.35}(a) as well as the relative difference with respect to the two LCFA spectra. The relative difference between the LCFA spectra is rather small being less than $0.5\%$ in the depicted energy regime. The relative difference of the accurate spectra, on the other hand, is about $3\%$, which is two times the value in Fig.~\ref{fig:spectra200-0.35}(b).

\begin{figure}[t]
  \begin{center}
  \includegraphics[width=0.49\textwidth]{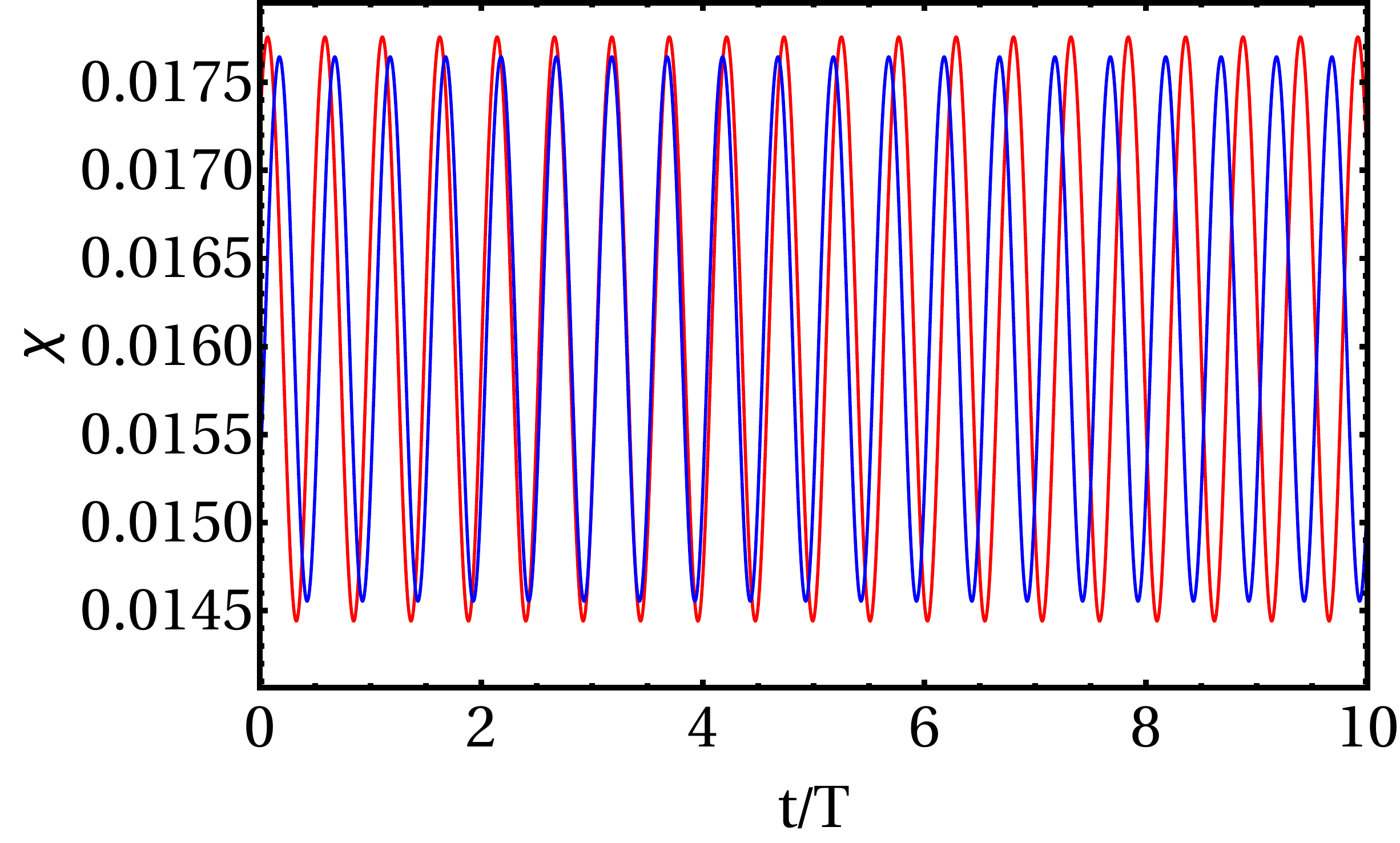}
  \caption{The quantum parameter $\chi$ for both co- and counter-rotating (red and blue) cases as a function of time. All the parameters are the same as in Fig.~\ref{fig:spectra200-0.35}.}
  \label{fig:chi}
  \end{center}
\end{figure}

The small difference between the LCFA spectra can be explained by looking at the quantum parameter $\chi$ in Fig.~\ref{fig:chi}. From this figure we can see that $\chi$ for both co- and counter-rotating setups has a similar average value and therefore gives almost the same LCFA results for both cases. The general behavior of $\chi$ can be estimated by  the well-known formula $\chi \approx |d\textbf{P}/d\tau|/m^2$, the small difference between co- and counter-rotating cases, however, are high order corrections that are not included in this estimation.    

The relative difference between the accurate spectra for two cases, in Fig.~\ref{fig:rela.diffs}(b), is only a few percent in the main part of the spectrum. This means that the first two differences (i) and (ii) mentioned before also did not cause a big deviation between the co- and counter-rotating cases. The reason is that when we calculate $|\mathcal{M}_{\mu}|^2=|\mathcal{M}_0|^2-|\mathcal{M}_1|^2-|\mathcal{M}_2|^2-|\mathcal{M}_3|^2$ in Eq.~\eqref{eq:dIdphi.final.fer}, the differences that appear in $\mathcal{M}_0$, $\mathcal{M}_2$, and $\mathcal{M}_3$ will contribute to the same degree in both cases. For example, in the co-rotating case, the crossing term is in $\mathcal{M}_3$ and the second term in $\mathcal{M}_2$ have the same sign in the final expression in Eq.~\eqref{eq:dIdphi.final.fer}. In the counter-rotating case, on the other hand, the crossing term moves to $\mathcal{M}_0$ and the second term in $\mathcal{M}_2$ changes the sign. Hence, in the end they will give a similar contribution to the co-rotating case. Moreover, the contribution is independent of the emitted photon energy as there is no obvious harmonic structure in the spectrum in this regime. This gives the linear increase of the relative difference with respect to the emitted photon energy in Fig.~\ref{fig:rela.diffs} since the prefactor in Eq.~\eqref{eq:dIdphi.final.fer} is proportional to $\omega'$.

\begin{figure}[t]
  \begin{center}
  \includegraphics[width=0.49\textwidth]{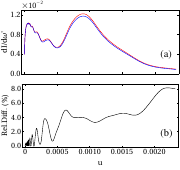}
  \caption{Panel (a): the radiation spectra as a function of $u$ for both co-rotating case (red) and counter-rotating case (blue). Panel (b): the relative difference between the two accurate spectra in panel (a).  Here, $\xi_1=20$ and $\xi_2=0.35$ and the average energy $\varepsilon=108m$, which corresponds to $\chi_1/\chi_2=0.5$.}
  \label{fig:spectra20-0.35}
  \end{center}
\end{figure}

For Regime II ($\chi_1 \sim \chi_2$), the two lasers will both play a role in the radiation spectra. In Fig.~\ref{fig:spectra20-0.35}(a), we show the spectra for both co- and counter-rotating cases with $\chi_1/\chi_2=0.5$. First of all, the two spectra again have the same shape. The second peak in the spectra is caused by the contributions from the $\xi_2$-laser as the position of this peak is roughly corresponding to the first harmonic of the $\xi_2$-laser with $u \approx 4\varepsilon^2\omega/(m^2\varepsilon-4\varepsilon^2\omega)=0.0094$. It is clearly shown that the deviation between the two spectra is more prominent. Because if $\chi_1 \sim \chi_2$ both lasers contribute to the radiation process therefore changing the rotating direction will cause a visible change in the spectrum. 

The relative difference between the co- and counter-rotating spectrum is displayed in Fig.~\ref{fig:spectra20-0.35}(b). As opposed to Regime I, the relative difference here is not linear in $u$ but oscillates. The reason is that with the present parameters the harmonic structure starts to be visible in the spectrum, and therefore difference (iv) will modify the position of the harmonics. For example, the position of the harmonics around $u=0.00035$ differs by $0.001m$ between the two cases. However, the main harmonics around $u=0.001$ are at the same place since this harmonic is mainly a consequence of $\xi_2$.

\begin{figure}[h]
  \begin{center}
  \includegraphics[width=0.49\textwidth]{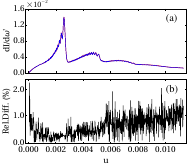}
  \caption{Panel (a): the radiation spectra as a function of $u$ for both co-rotating case (red) and counter-rotating case (blue). Panel (b): the relative difference between the two accurate spectra in panel (a).  Here, $\xi_1=20$ and $\xi_2=1$ and the average energy $\varepsilon=447m$, which corresponds to $\chi_1/\chi_2=0.01$.}
  \label{fig:spectra20-1}
  \end{center}
\end{figure}

For Regime III ($\chi_1 \ll \chi_2$), the spectrum will be dominated by $\xi_2$ laser now. From Fig.~\ref{fig:spectra20-1}(a), we can see the main structure of the spectra consists of the harmonics created by $\xi_2$. The additional fast oscillation is the modification from $\xi_1$. Like in the previous two regimes, the spectra for both cases are the same as the relative difference between them in panel (b) is less than $1\%$. This is because in the single laser case, the change of the sense of rotation will not affect the radiation process. Please note that there are two curves on top of each other in panel (a).  

The fast oscillation in the relative difference is due to the harmonics being shifted by different $s_2$ for co- and counter-rotating cases [difference (iv)]. However, the main peak of the spectrum is at the same position because it orignates mainly from the $\xi_2$-laser. Moreover, the average relative difference also increases approximately linearly for a large photon energy, which is due to the spectrum behaving alike the predictions of the LCFA at large emitted energies.

Comparing the relative difference in all three regimes I, II, and III, it shows that the relative difference is larger in Regime II but smaller in both Regime I and III. The reason is that when $\chi_1 \ll \chi_2$ or $\chi_1 \gg \chi_2$ only one of the lasers dominates the radiation and therefore the change of the rotation direction will not affect the spectrum noteably as in the single laser case. For $\chi_1 \sim \chi_2$, on the other hand, both lasers contribute to the radiation and changing the laser's sense of rotation will induce a more significant alteration of the spectrum compared to the previous two cases. However, in all the regimes, the spectra for both cases have the same shape, which is due to that the characteristic time scale of the electron in both cases is the same. To understand this similiarity, we can go back to the orginal formula for the Baier-Katkov integral in Eq.~\eqref{eq:T.def.t}, where the velocity of the electron plays a majar role. However, the velocities for both cases are simular as the different crossing term in Eq.~\eqref{eq:four.momen.final} is of second order in amplitude. More intuitively, we know that for an ultrarelativistic electron in a strong background field the emission is mainly along the propagation direction within the $1/\gamma$ cone and the shape of the spectrum is determined by the electron dynamics in this region, which can be characterized by a time scale defined as $t_c \coloneqq |\dot{\textbf{v}}|/|\ddot{\textbf{v}}|$ with $\textbf{v}$ being the electron's velocity in the fields. Based on the momentum in Eq.~\eqref{eq:four.momen.final}, $t_c$ is similar for both cases despite that the crossing term moves from $P_z$ in the co-rotating case to $\varepsilon$ in the counter-rotating one.
 
\section{Conclusions}
\label{sec:conc}
In this paper the dynamics of an ultrarelativistic electron in counterpropagating laser beams with variable sense of rotation have been explored. The classical momentum and trajectory are analytically derived assuming that the particle's averaged energy is the dominant factor and the transverse momentum is small compared to the total energy. The difference between the classical momentum for both co- and counter-rotating laser beam arrangements has been investigated. The main difference appeared to be the crossing term related to both laser beams that moves from $P_z$ in the co-rotating case to the total energy $\varepsilon$ in the counter-rotating case, see Eqs.~\eqref{eq:momen.Pz} and \eqref{eq:en.final}. This means that for a vanishing initial transverse momentum the electron in a counter-rotating setup has a constant velocity along $z$-direction. And if we change to the frame that goes together with the electron along $z$-direction, the electron's trajectory in this setup is almost the same as in a two-color rotating electric field confioguration, which is a widely used model for analyzing the so-called dynamically assisted Schwinger effect \cite{Otto_2015,Schneider_2016,Torgrimsson_2017,Torgrimsson_2018,Linder_2015,Aleksandrov_2018,DiPiazza_2009,Dunne_2008}. Therefore, the counter-rotating setup can provide a mapping of this simple model for the whole space region rather than only a small vicinity around the antinode of a standing wave, where the rotating electric field can be realized. Moreover, a comparison with the full numerical solution was carried out resulting in a good agreement and validation of our analytical solution within the given conditions. 

The closed formula of the analytical approximation for the classical electron dynamics allows one to calculate the rates of the quantum processes in strong CPW background fields employing the Baier-Katkov semi-classical operator method. In this formalism, while the electron dynamics in the background classical fields are accounted for quasiclassically, the photon emission is treated quantum mechanically, fully taking into account the quantum recoil of the emitted photon. The Baier-Katkov integrals were analytically solved yielding closed formulas in terms of various Bessel functions. Different expressions are obtained for the co-rotating and counter-rotating cases, respectively. The results are employed to compare the co-rotating and counter-rotating cases in detail. We have observed that even though the classical dynamics show a qualitative difference between the two cases the emitted spectra all have the same shape for two cases in different parameter regimes with a relative difference of only a few percent. To understand these deviations, we have analyzed the arguments of the Bessel functions in the matrix elements in detail. The influence of the rotational direction is prominent only when the radiation process is dominated by both laser beams.

\acknowledgments

L.~A. would like to thank the group of Professor C. H. Keitel for the nice hospitality at MPIK in Heidelberg. The authors would like to thank Dr. E. Raicher and Dr. K.~Z. Hatsagortsyan for helpful discussions during the project, and also for reading and useful comments on the manuscript.

\bibliography{lcfa}

\begin{thebibliography}{68}%
\makeatletter
\providecommand \@ifxundefined [1]{%
 \@ifx{#1\undefined}
}%
\providecommand \@ifnum [1]{%
 \ifnum #1\expandafter \@firstoftwo
 \else \expandafter \@secondoftwo
 \fi
}%
\providecommand \@ifx [1]{%
 \ifx #1\expandafter \@firstoftwo
 \else \expandafter \@secondoftwo
 \fi
}%
\providecommand \natexlab [1]{#1}%
\providecommand \enquote  [1]{``#1''}%
\providecommand \bibnamefont  [1]{#1}%
\providecommand \bibfnamefont [1]{#1}%
\providecommand \citenamefont [1]{#1}%
\providecommand \href@noop [0]{\@secondoftwo}%
\providecommand \href [0]{\begingroup \@sanitize@url \@href}%
\providecommand \@href[1]{\@@startlink{#1}\@@href}%
\providecommand \@@href[1]{\endgroup#1\@@endlink}%
\providecommand \@sanitize@url [0]{\catcode `\\12\catcode `\$12\catcode
  `\&12\catcode `\#12\catcode `\^12\catcode `\_12\catcode `\%12\relax}%
\providecommand \@@startlink[1]{}%
\providecommand \@@endlink[0]{}%
\providecommand \url  [0]{\begingroup\@sanitize@url \@url }%
\providecommand \@url [1]{\endgroup\@href {#1}{\urlprefix }}%
\providecommand \urlprefix  [0]{URL }%
\providecommand \Eprint [0]{\href }%
\providecommand \doibase [0]{http://dx.doi.org/}%
\providecommand \selectlanguage [0]{\@gobble}%
\providecommand \bibinfo  [0]{\@secondoftwo}%
\providecommand \bibfield  [0]{\@secondoftwo}%
\providecommand \translation [1]{[#1]}%
\providecommand \BibitemOpen [0]{}%
\providecommand \bibitemStop [0]{}%
\providecommand \bibitemNoStop [0]{.\EOS\space}%
\providecommand \EOS [0]{\spacefactor3000\relax}%
\providecommand \BibitemShut  [1]{\csname bibitem#1\endcsname}%
\let\auto@bib@innerbib\@empty
\bibitem [{\citenamefont {{The Vulcan facility}}()}]{Vulcan1}%
  \BibitemOpen
  \bibfield  {author} {\bibinfo {author} {\bibnamefont {{The Vulcan
  facility}}},\ }\href@noop {} {}\bibinfo {howpublished}
  {\url{https://www.clf.stfc.ac.uk/Pages/ Vulcan-laser.aspx}}\BibitemShut
  {NoStop}%
\bibitem [{\citenamefont {Yoon}\ \emph {et~al.}(2019)\citenamefont {Yoon},
  \citenamefont {Jeon}, \citenamefont {Shin}, \citenamefont {Lee},
  \citenamefont {Lee}, \citenamefont {Choi}, \citenamefont {Kim}, \citenamefont
  {Sung}, ,\ and\ \citenamefont {Nam}}]{Yoon_2019}%
  \BibitemOpen
  \bibfield  {author} {\bibinfo {author} {\bibfnamefont {J.W.}\ \bibnamefont
  {Yoon}}, \bibinfo {author} {\bibfnamefont {C.}~\bibnamefont {Jeon}}, \bibinfo
  {author} {\bibfnamefont {J.}~\bibnamefont {Shin}}, \bibinfo {author}
  {\bibfnamefont {S.K.}\ \bibnamefont {Lee}}, \bibinfo {author} {\bibfnamefont
  {H.W.}\ \bibnamefont {Lee}}, \bibinfo {author} {\bibfnamefont {I.W.}\
  \bibnamefont {Choi}}, \bibinfo {author} {\bibfnamefont {H.T.}\ \bibnamefont
  {Kim}}, \bibinfo {author} {\bibfnamefont {J.H.}\ \bibnamefont {Sung}}, , \
  and\ \bibinfo {author} {\bibfnamefont {C.H.}\ \bibnamefont {Nam}},\
  }\href@noop {} {\bibfield  {journal} {\bibinfo  {journal} {Opt. Express}\
  }\textbf {\bibinfo {volume} {27}},\ \bibinfo {pages} {20412} (\bibinfo {year}
  {2019})}\BibitemShut {NoStop}%
\bibitem [{\citenamefont {{The Extreme Light Infrastructure (ELI)}}()}]{ELI}%
  \BibitemOpen
  \bibfield  {author} {\bibinfo {author} {\bibnamefont {{The Extreme Light
  Infrastructure (ELI)}}},\ }\href@noop {} {}\bibinfo {howpublished}
  {\url{http://www.eli-laser.eu/}}\BibitemShut {NoStop}%
\bibitem [{\citenamefont {{Exawatt Center for Extreme Light Stidies
  (XCELS)}}()}]{XCELS}%
  \BibitemOpen
  \bibfield  {author} {\bibinfo {author} {\bibnamefont {{Exawatt Center for
  Extreme Light Stidies (XCELS)}}},\ }\href@noop {} {}\bibinfo {howpublished}
  {\url{http://www.xcels.iapras.ru/}}\BibitemShut {NoStop}%
\bibitem [{\citenamefont {Marklund}\ and\ \citenamefont
  {Shukla}(2006)}]{Marklund_2006}%
  \BibitemOpen
  \bibfield  {author} {\bibinfo {author} {\bibfnamefont {M.}~\bibnamefont
  {Marklund}}\ and\ \bibinfo {author} {\bibfnamefont {P.~K.}\ \bibnamefont
  {Shukla}},\ }\bibfield  {title} {\enquote {\bibinfo {title} {Nonlinear
  collective effects in photon-photon and photon-plasma interactions},}\
  }\href@noop {} {\bibfield  {journal} {\bibinfo  {journal} {Rev. Mod. Phys.}\
  }\textbf {\bibinfo {volume} {78}},\ \bibinfo {pages} {591} (\bibinfo {year}
  {2006})}\BibitemShut {NoStop}%
\bibitem [{\citenamefont {{Di Piazza}}\ \emph {et~al.}(2012)\citenamefont {{Di
  Piazza}}, \citenamefont {M\"uller}, \citenamefont {Hatsagortsyan},\ and\
  \citenamefont {Keitel}}]{RMP_2012}%
  \BibitemOpen
  \bibfield  {author} {\bibinfo {author} {\bibfnamefont {A.}~\bibnamefont {{Di
  Piazza}}}, \bibinfo {author} {\bibfnamefont {C.}~\bibnamefont {M\"uller}},
  \bibinfo {author} {\bibfnamefont {K.~Z.}\ \bibnamefont {Hatsagortsyan}}, \
  and\ \bibinfo {author} {\bibfnamefont {C.~H.}\ \bibnamefont {Keitel}},\
  }\bibfield  {title} {\enquote {\bibinfo {title} {Extremely high-intensity
  laser interactions with fundamental quantum systems},}\ }\href@noop {}
  {\bibfield  {journal} {\bibinfo  {journal} {Rev. Mod. Phys.}\ }\textbf
  {\bibinfo {volume} {84}},\ \bibinfo {pages} {1177} (\bibinfo {year}
  {2012})}\BibitemShut {NoStop}%
\bibitem [{\citenamefont {Heinzl}(2012)}]{Heinzl_2012}%
  \BibitemOpen
  \bibfield  {author} {\bibinfo {author} {\bibfnamefont {T.}~\bibnamefont
  {Heinzl}},\ }\bibfield  {title} {\enquote {\bibinfo {title} {Strong-field qed
  and high-power lasers},}\ }\href@noop {} {\bibfield  {journal} {\bibinfo
  {journal} {Int. J. Mod. Phys. A}\ }\textbf {\bibinfo {volume} {27}},\
  \bibinfo {pages} {1260010} (\bibinfo {year} {2012})}\BibitemShut {NoStop}%
\bibitem [{\citenamefont {Dunne}(2014)}]{Dunne2014}%
  \BibitemOpen
  \bibfield  {author} {\bibinfo {author} {\bibfnamefont {G.V.}\ \bibnamefont
  {Dunne}},\ }\href@noop {} {\bibfield  {journal} {\bibinfo  {journal} {Eur.
  Phys. J. Spec. Top.}\ }\textbf {\bibinfo {volume} {223}},\ \bibinfo {pages}
  {1055} (\bibinfo {year} {2014})}\BibitemShut {NoStop}%
\bibitem [{\citenamefont {Turcu}\ \emph {et~al.}(2019)\citenamefont {Turcu},
  \citenamefont {Shen}, \citenamefont {Neely}, \citenamefont {Sarri},
  \citenamefont {Tanaka}, \citenamefont {McKenna}, \citenamefont {Mangles},
  \citenamefont {Yu}, \citenamefont {Luo}, \citenamefont {Zhu},\ and\
  \citenamefont {et~al.}}]{Turcu_2019}%
  \BibitemOpen
  \bibfield  {author} {\bibinfo {author} {\bibfnamefont {I.~C.~E.}\
  \bibnamefont {Turcu}}, \bibinfo {author} {\bibfnamefont {B.}~\bibnamefont
  {Shen}}, \bibinfo {author} {\bibfnamefont {D.}~\bibnamefont {Neely}},
  \bibinfo {author} {\bibfnamefont {G.}~\bibnamefont {Sarri}}, \bibinfo
  {author} {\bibfnamefont {K.~A.}\ \bibnamefont {Tanaka}}, \bibinfo {author}
  {\bibfnamefont {P.}~\bibnamefont {McKenna}}, \bibinfo {author} {\bibfnamefont
  {S.~P.~D.}\ \bibnamefont {Mangles}}, \bibinfo {author} {\bibfnamefont
  {T.-P.}\ \bibnamefont {Yu}}, \bibinfo {author} {\bibfnamefont
  {W.}~\bibnamefont {Luo}}, \bibinfo {author} {\bibfnamefont {X.-L.}\
  \bibnamefont {Zhu}}, \ and\ \bibinfo {author} {\bibnamefont {et~al.}},\
  }\bibfield  {title} {\enquote {\bibinfo {title} {Quantum electrodynamics
  experiments with colliding petawatt laser pulses},}\ }\href@noop {}
  {\bibfield  {journal} {\bibinfo  {journal} {High Power Laser Sci. Eng.}\
  }\textbf {\bibinfo {volume} {7}},\ \bibinfo {pages} {e10} (\bibinfo {year}
  {2019})}\BibitemShut {NoStop}%
\bibitem [{\citenamefont {Cole}\ \emph {et~al.}(2018)\citenamefont {Cole},
  \citenamefont {Behm}, \citenamefont {Gerstmayr}, \citenamefont {Blackburn},
  \citenamefont {Wood}, \citenamefont {Baird}, \citenamefont {Duff},
  \citenamefont {Harvey}, \citenamefont {Ilderton}, \citenamefont {Joglekar},
  \citenamefont {Krushelnick}, \citenamefont {Kuschel}, \citenamefont
  {Marklund}, \citenamefont {McKenna}, \citenamefont {Murphy}, \citenamefont
  {Poder}, \citenamefont {Ridgers}, \citenamefont {Samarin}, \citenamefont
  {Sarri}, \citenamefont {Symes}, \citenamefont {Thomas}, \citenamefont
  {Warwick}, \citenamefont {Zepf}, \citenamefont {Najmudin},\ and\
  \citenamefont {Mangles}}]{Cole_2018}%
  \BibitemOpen
  \bibfield  {author} {\bibinfo {author} {\bibfnamefont {J.~M.}\ \bibnamefont
  {Cole}}, \bibinfo {author} {\bibfnamefont {K.~T.}\ \bibnamefont {Behm}},
  \bibinfo {author} {\bibfnamefont {E.}~\bibnamefont {Gerstmayr}}, \bibinfo
  {author} {\bibfnamefont {T.~G.}\ \bibnamefont {Blackburn}}, \bibinfo {author}
  {\bibfnamefont {J.~C.}\ \bibnamefont {Wood}}, \bibinfo {author}
  {\bibfnamefont {C.~D.}\ \bibnamefont {Baird}}, \bibinfo {author}
  {\bibfnamefont {M.~J.}\ \bibnamefont {Duff}}, \bibinfo {author}
  {\bibfnamefont {C.}~\bibnamefont {Harvey}}, \bibinfo {author} {\bibfnamefont
  {A.}~\bibnamefont {Ilderton}}, \bibinfo {author} {\bibfnamefont {A.~S.}\
  \bibnamefont {Joglekar}}, \bibinfo {author} {\bibfnamefont {K.}~\bibnamefont
  {Krushelnick}}, \bibinfo {author} {\bibfnamefont {S.}~\bibnamefont
  {Kuschel}}, \bibinfo {author} {\bibfnamefont {M.}~\bibnamefont {Marklund}},
  \bibinfo {author} {\bibfnamefont {P.}~\bibnamefont {McKenna}}, \bibinfo
  {author} {\bibfnamefont {C.~D.}\ \bibnamefont {Murphy}}, \bibinfo {author}
  {\bibfnamefont {K.}~\bibnamefont {Poder}}, \bibinfo {author} {\bibfnamefont
  {C.~P.}\ \bibnamefont {Ridgers}}, \bibinfo {author} {\bibfnamefont {G.~M.}\
  \bibnamefont {Samarin}}, \bibinfo {author} {\bibfnamefont {G.}~\bibnamefont
  {Sarri}}, \bibinfo {author} {\bibfnamefont {D.~R.}\ \bibnamefont {Symes}},
  \bibinfo {author} {\bibfnamefont {A.~G.~R.}\ \bibnamefont {Thomas}}, \bibinfo
  {author} {\bibfnamefont {J.}~\bibnamefont {Warwick}}, \bibinfo {author}
  {\bibfnamefont {M.}~\bibnamefont {Zepf}}, \bibinfo {author} {\bibfnamefont
  {Z.}~\bibnamefont {Najmudin}}, \ and\ \bibinfo {author} {\bibfnamefont
  {S.~P.~D.}\ \bibnamefont {Mangles}},\ }\bibfield  {title} {\enquote {\bibinfo
  {title} {Experimental evidence of radiation reaction in the collision of a
  high-intensity laser pulse with a laser-wakefield accelerated electron
  beam},}\ }\href@noop {} {\bibfield  {journal} {\bibinfo  {journal} {Phys.
  Rev. X}\ }\textbf {\bibinfo {volume} {8}},\ \bibinfo {pages} {011020}
  (\bibinfo {year} {2018})}\BibitemShut {NoStop}%
\bibitem [{\citenamefont {Poder}\ \emph {et~al.}(2018)\citenamefont {Poder},
  \citenamefont {Tamburini}, \citenamefont {Sarri}, \citenamefont {Di~Piazza},
  \citenamefont {Kuschel}, \citenamefont {Baird}, \citenamefont {Behm},
  \citenamefont {Bohlen}, \citenamefont {Cole}, \citenamefont {Corvan},
  \citenamefont {Duff}, \citenamefont {Gerstmayr}, \citenamefont {Keitel},
  \citenamefont {Krushelnick}, \citenamefont {Mangles}, \citenamefont
  {McKenna}, \citenamefont {Murphy}, \citenamefont {Najmudin}, \citenamefont
  {Ridgers}, \citenamefont {Samarin}, \citenamefont {Symes}, \citenamefont
  {Thomas}, \citenamefont {Warwick},\ and\ \citenamefont {Zepf}}]{Poder_2018}%
  \BibitemOpen
  \bibfield  {author} {\bibinfo {author} {\bibfnamefont {K.}~\bibnamefont
  {Poder}}, \bibinfo {author} {\bibfnamefont {M.}~\bibnamefont {Tamburini}},
  \bibinfo {author} {\bibfnamefont {G.}~\bibnamefont {Sarri}}, \bibinfo
  {author} {\bibfnamefont {A.}~\bibnamefont {Di~Piazza}}, \bibinfo {author}
  {\bibfnamefont {S.}~\bibnamefont {Kuschel}}, \bibinfo {author} {\bibfnamefont
  {C.~D.}\ \bibnamefont {Baird}}, \bibinfo {author} {\bibfnamefont
  {K.}~\bibnamefont {Behm}}, \bibinfo {author} {\bibfnamefont {S.}~\bibnamefont
  {Bohlen}}, \bibinfo {author} {\bibfnamefont {J.~M.}\ \bibnamefont {Cole}},
  \bibinfo {author} {\bibfnamefont {D.~J.}\ \bibnamefont {Corvan}}, \bibinfo
  {author} {\bibfnamefont {M.}~\bibnamefont {Duff}}, \bibinfo {author}
  {\bibfnamefont {E.}~\bibnamefont {Gerstmayr}}, \bibinfo {author}
  {\bibfnamefont {C.~H.}\ \bibnamefont {Keitel}}, \bibinfo {author}
  {\bibfnamefont {K.}~\bibnamefont {Krushelnick}}, \bibinfo {author}
  {\bibfnamefont {S.~P.~D.}\ \bibnamefont {Mangles}}, \bibinfo {author}
  {\bibfnamefont {P.}~\bibnamefont {McKenna}}, \bibinfo {author} {\bibfnamefont
  {C.~D.}\ \bibnamefont {Murphy}}, \bibinfo {author} {\bibfnamefont
  {Z.}~\bibnamefont {Najmudin}}, \bibinfo {author} {\bibfnamefont {C.~P.}\
  \bibnamefont {Ridgers}}, \bibinfo {author} {\bibfnamefont {G.~M.}\
  \bibnamefont {Samarin}}, \bibinfo {author} {\bibfnamefont {D.~R.}\
  \bibnamefont {Symes}}, \bibinfo {author} {\bibfnamefont {A.~G.~R.}\
  \bibnamefont {Thomas}}, \bibinfo {author} {\bibfnamefont {J.}~\bibnamefont
  {Warwick}}, \ and\ \bibinfo {author} {\bibfnamefont {M.}~\bibnamefont
  {Zepf}},\ }\bibfield  {title} {\enquote {\bibinfo {title} {Experimental
  signatures of the quantum nature of radiation reaction in the field of an
  ultraintense laser},}\ }\href@noop {} {\bibfield  {journal} {\bibinfo
  {journal} {Phys. Rev. X}\ }\textbf {\bibinfo {volume} {8}},\ \bibinfo {pages}
  {031004} (\bibinfo {year} {2018})}\BibitemShut {NoStop}%
\bibitem [{\citenamefont {Furry}(1951)}]{Furry_1951}%
  \BibitemOpen
  \bibfield  {author} {\bibinfo {author} {\bibfnamefont {W.~H.}\ \bibnamefont
  {Furry}},\ }\bibfield  {title} {\enquote {\bibinfo {title} {On bound states
  and scattering in positron theory},}\ }\href@noop {} {\bibfield  {journal}
  {\bibinfo  {journal} {Phys. Rev.}\ }\textbf {\bibinfo {volume} {81}},\
  \bibinfo {pages} {115} (\bibinfo {year} {1951})}\BibitemShut {NoStop}%
\bibitem [{\citenamefont {Bagrov}\ and\ \citenamefont
  {Gitman}(1990)}]{Bagrov_1990}%
  \BibitemOpen
  \bibfield  {author} {\bibinfo {author} {\bibfnamefont
  {Vladislav~Gavrilovich}\ \bibnamefont {Bagrov}}\ and\ \bibinfo {author}
  {\bibfnamefont {D}~\bibnamefont {Gitman}},\ }\href@noop {} {\emph {\bibinfo
  {title} {Exact solutions of relativistic wave equations}}},\ Vol.~\bibinfo
  {volume} {39}\ (\bibinfo  {publisher} {Springer Science \& Business Media},\
  \bibinfo {year} {1990})\BibitemShut {NoStop}%
\bibitem [{\citenamefont {{Di Piazza}}\ \emph {et~al.}(2018)\citenamefont {{Di
  Piazza}}, \citenamefont {Tamburini}, \citenamefont {Meuren},\ and\
  \citenamefont {Keitel}}]{DiPiazza_2018}%
  \BibitemOpen
  \bibfield  {author} {\bibinfo {author} {\bibfnamefont {A}~\bibnamefont {{Di
  Piazza}}}, \bibinfo {author} {\bibfnamefont {M}~\bibnamefont {Tamburini}},
  \bibinfo {author} {\bibfnamefont {S}~\bibnamefont {Meuren}}, \ and\ \bibinfo
  {author} {\bibfnamefont {C~H}\ \bibnamefont {Keitel}},\ }\bibfield  {title}
  {\enquote {\bibinfo {title} {{Implementing nonlinear Compton scattering
  beyond the local-constant-field approximation}},}\ }\href@noop {} {\bibfield
  {journal} {\bibinfo  {journal} {Phys. Rev. A}\ }\textbf {\bibinfo {volume}
  {98}},\ \bibinfo {pages} {012134} (\bibinfo {year} {2018})}\BibitemShut
  {NoStop}%
\bibitem [{\citenamefont {{Di Piazza}}\ \emph {et~al.}(2019)\citenamefont {{Di
  Piazza}}, \citenamefont {Tamburini}, \citenamefont {Meuren},\ and\
  \citenamefont {Keitel}}]{DiPiazza_2019}%
  \BibitemOpen
  \bibfield  {author} {\bibinfo {author} {\bibfnamefont {A}~\bibnamefont {{Di
  Piazza}}}, \bibinfo {author} {\bibfnamefont {M}~\bibnamefont {Tamburini}},
  \bibinfo {author} {\bibfnamefont {S}~\bibnamefont {Meuren}}, \ and\ \bibinfo
  {author} {\bibfnamefont {C~H}\ \bibnamefont {Keitel}},\ }\bibfield  {title}
  {\enquote {\bibinfo {title} {{Improved local-constant-field approximation for
  strong-field QED codes}},}\ }\href@noop {} {\bibfield  {journal} {\bibinfo
  {journal} {Phys. Rev. A}\ }\textbf {\bibinfo {volume} {99}},\ \bibinfo
  {pages} {022125} (\bibinfo {year} {2019})}\BibitemShut {NoStop}%
\bibitem [{\citenamefont {Ilderton}\ \emph
  {et~al.}(2019{\natexlab{a}})\citenamefont {Ilderton}, \citenamefont {King},\
  and\ \citenamefont {Seipt}}]{Ilderton_2019a}%
  \BibitemOpen
  \bibfield  {author} {\bibinfo {author} {\bibfnamefont {A.}~\bibnamefont
  {Ilderton}}, \bibinfo {author} {\bibfnamefont {B.}~\bibnamefont {King}}, \
  and\ \bibinfo {author} {\bibfnamefont {D.}~\bibnamefont {Seipt}},\ }\bibfield
   {title} {\enquote {\bibinfo {title} {{Extended locally constant field
  approximation for nonlinear Compton scattering}},}\ }\href@noop {} {\bibfield
   {journal} {\bibinfo  {journal} {Phys. Rev. A}\ }\textbf {\bibinfo {volume}
  {99}},\ \bibinfo {pages} {042121} (\bibinfo {year}
  {2019}{\natexlab{a}})}\BibitemShut {NoStop}%
\bibitem [{\citenamefont {Ilderton}\ \emph
  {et~al.}(2019{\natexlab{b}})\citenamefont {Ilderton}, \citenamefont {King},\
  and\ \citenamefont {MacLeod}}]{Ilderton_2019b}%
  \BibitemOpen
  \bibfield  {author} {\bibinfo {author} {\bibfnamefont {A.}~\bibnamefont
  {Ilderton}}, \bibinfo {author} {\bibfnamefont {B.}~\bibnamefont {King}}, \
  and\ \bibinfo {author} {\bibfnamefont {A.~J.}\ \bibnamefont {MacLeod}},\
  }\bibfield  {title} {\enquote {\bibinfo {title} {Absorption cross section in
  an intense plane wave background},}\ }\href@noop {} {\bibfield  {journal}
  {\bibinfo  {journal} {Phys. Rev. D}\ }\textbf {\bibinfo {volume} {100}},\
  \bibinfo {pages} {076002} (\bibinfo {year} {2019}{\natexlab{b}})}\BibitemShut
  {NoStop}%
\bibitem [{\citenamefont {Podszus}\ and\ \citenamefont
  {Di~Piazza}(2019)}]{Podszus_2019}%
  \BibitemOpen
  \bibfield  {author} {\bibinfo {author} {\bibfnamefont {T.}~\bibnamefont
  {Podszus}}\ and\ \bibinfo {author} {\bibfnamefont {A.}~\bibnamefont
  {Di~Piazza}},\ }\bibfield  {title} {\enquote {\bibinfo {title} {High-energy
  behavior of strong-field qed in an intense plane wave},}\ }\href@noop {}
  {\bibfield  {journal} {\bibinfo  {journal} {Phys. Rev. D}\ }\textbf {\bibinfo
  {volume} {99}},\ \bibinfo {pages} {076004} (\bibinfo {year}
  {2019})}\BibitemShut {NoStop}%
\bibitem [{\citenamefont {Lv}\ \emph {et~al.}(2021{\natexlab{a}})\citenamefont
  {Lv}, \citenamefont {Raicher}, \citenamefont {Keitel},\ and\ \citenamefont
  {Hatsagortsyan}}]{Lv_2021a}%
  \BibitemOpen
  \bibfield  {author} {\bibinfo {author} {\bibfnamefont {Q.~Z.}\ \bibnamefont
  {Lv}}, \bibinfo {author} {\bibfnamefont {E.}~\bibnamefont {Raicher}},
  \bibinfo {author} {\bibfnamefont {C.~H.}\ \bibnamefont {Keitel}}, \ and\
  \bibinfo {author} {\bibfnamefont {K.~Z.}\ \bibnamefont {Hatsagortsyan}},\
  }\bibfield  {title} {\enquote {\bibinfo {title} {Anomalous violation of the
  local constant field approximation in colliding laser beams},}\ }\href@noop
  {} {\bibfield  {journal} {\bibinfo  {journal} {Physical Review Research}\
  }\textbf {\bibinfo {volume} {3}},\ \bibinfo {pages} {013214} (\bibinfo {year}
  {2021}{\natexlab{a}})}\BibitemShut {NoStop}%
\bibitem [{\citenamefont {Popov}\ \emph {et~al.}(1997)\citenamefont {Popov},
  \citenamefont {Mur},\ and\ \citenamefont {Karnakov}}]{Popov_1997}%
  \BibitemOpen
  \bibfield  {author} {\bibinfo {author} {\bibfnamefont {V.}~\bibnamefont
  {Popov}}, \bibinfo {author} {\bibfnamefont {V.}~\bibnamefont {Mur}}, \ and\
  \bibinfo {author} {\bibfnamefont {B.}~\bibnamefont {Karnakov}},\ }\href@noop
  {} {\bibfield  {journal} {\bibinfo  {journal} {JETP Letters}\ }\textbf
  {\bibinfo {volume} {66}},\ \bibinfo {pages} {229} (\bibinfo {year}
  {1997})}\BibitemShut {NoStop}%
\bibitem [{\citenamefont {Gersten}\ and\ \citenamefont
  {Mittleman}(1975)}]{Gersten_1975}%
  \BibitemOpen
  \bibfield  {author} {\bibinfo {author} {\bibfnamefont {Joel~I.}\ \bibnamefont
  {Gersten}}\ and\ \bibinfo {author} {\bibfnamefont {Marvin~H.}\ \bibnamefont
  {Mittleman}},\ }\bibfield  {title} {\enquote {\bibinfo {title} {Eikonal
  theory of charged-particle scattering in the presence of a strong
  electromagnetic wave},}\ }\href@noop {} {\bibfield  {journal} {\bibinfo
  {journal} {Phys. Rev. A}\ }\textbf {\bibinfo {volume} {12}},\ \bibinfo
  {pages} {1840--1845} (\bibinfo {year} {1975})}\BibitemShut {NoStop}%
\bibitem [{\citenamefont {Mocken}\ \emph {et~al.}(2010)\citenamefont {Mocken},
  \citenamefont {Ruf}, \citenamefont {M\"uller},\ and\ \citenamefont
  {Keitel}}]{Mocken_2010}%
  \BibitemOpen
  \bibfield  {author} {\bibinfo {author} {\bibfnamefont {G.~R.}\ \bibnamefont
  {Mocken}}, \bibinfo {author} {\bibfnamefont {M.}~\bibnamefont {Ruf}},
  \bibinfo {author} {\bibfnamefont {C.}~\bibnamefont {M\"uller}}, \ and\
  \bibinfo {author} {\bibfnamefont {C.~H.}\ \bibnamefont {Keitel}},\ }\bibfield
   {title} {\enquote {\bibinfo {title} {Nonperturbative multiphoton
  electron-positron--pair creation in laser fields},}\ }\href@noop {}
  {\bibfield  {journal} {\bibinfo  {journal} {Phys. Rev. A}\ }\textbf {\bibinfo
  {volume} {81}},\ \bibinfo {pages} {022122} (\bibinfo {year}
  {2010})}\BibitemShut {NoStop}%
\bibitem [{\citenamefont {{Di~Piazza}}(2014)}]{DiPiazza_2014}%
  \BibitemOpen
  \bibfield  {author} {\bibinfo {author} {\bibfnamefont {A.}~\bibnamefont
  {{Di~Piazza}}},\ }\bibfield  {title} {\enquote {\bibinfo {title}
  {Ultrarelativistic electron states in a general background electromagnetic
  field},}\ }\href@noop {} {\bibfield  {journal} {\bibinfo  {journal} {Phys.
  Rev. Lett.}\ }\textbf {\bibinfo {volume} {113}},\ \bibinfo {pages} {040402}
  (\bibinfo {year} {2014})}\BibitemShut {NoStop}%
\bibitem [{\citenamefont {Di~Piazza}(2015)}]{DiPiazza_2015}%
  \BibitemOpen
  \bibfield  {author} {\bibinfo {author} {\bibfnamefont {A.}~\bibnamefont
  {Di~Piazza}},\ }\bibfield  {title} {\enquote {\bibinfo {title} {Analytical
  tools for investigating strong-field qed processes in tightly focused laser
  fields},}\ }\href@noop {} {\bibfield  {journal} {\bibinfo  {journal} {Phys.
  Rev. A}\ }\textbf {\bibinfo {volume} {91}},\ \bibinfo {pages} {042118}
  (\bibinfo {year} {2015})}\BibitemShut {NoStop}%
\bibitem [{\citenamefont {Di~Piazza}(2016)}]{DiPiazza_2016}%
  \BibitemOpen
  \bibfield  {author} {\bibinfo {author} {\bibfnamefont {A.}~\bibnamefont
  {Di~Piazza}},\ }\bibfield  {title} {\enquote {\bibinfo {title} {Nonlinear
  breit-wheeler pair production in a tightly focused laser beam},}\ }\href@noop
  {} {\bibfield  {journal} {\bibinfo  {journal} {Phys. Rev. Lett.}\ }\textbf
  {\bibinfo {volume} {117}},\ \bibinfo {pages} {213201} (\bibinfo {year}
  {2016})}\BibitemShut {NoStop}%
\bibitem [{\citenamefont {Di~Piazza}(2017)}]{DiPiazza_2017}%
  \BibitemOpen
  \bibfield  {author} {\bibinfo {author} {\bibfnamefont {A.}~\bibnamefont
  {Di~Piazza}},\ }\bibfield  {title} {\enquote {\bibinfo {title} {First-order
  strong-field qed processes in a tightly focused laser beam},}\ }\href@noop {}
  {\bibfield  {journal} {\bibinfo  {journal} {Phys. Rev. A}\ }\textbf {\bibinfo
  {volume} {95}},\ \bibinfo {pages} {032121} (\bibinfo {year}
  {2017})}\BibitemShut {NoStop}%
\bibitem [{\citenamefont {Baier}\ and\ \citenamefont
  {Katkov}(1968)}]{Katkov_1968}%
  \BibitemOpen
  \bibfield  {author} {\bibinfo {author} {\bibfnamefont {V.~N.}\ \bibnamefont
  {Baier}}\ and\ \bibinfo {author} {\bibfnamefont {V.~M.}\ \bibnamefont
  {Katkov}},\ }\href@noop {} {\bibfield  {journal} {\bibinfo  {journal} {Sov.
  Phys. JETP}\ }\textbf {\bibinfo {volume} {26}},\ \bibinfo {pages} {854}
  (\bibinfo {year} {1968})}\BibitemShut {NoStop}%
\bibitem [{\citenamefont {Baier}\ \emph {et~al.}(1994)\citenamefont {Baier},
  \citenamefont {Katkov},\ and\ \citenamefont {Strakhovenko}}]{Baier_b_1994}%
  \BibitemOpen
  \bibfield  {author} {\bibinfo {author} {\bibfnamefont {V.~N.}\ \bibnamefont
  {Baier}}, \bibinfo {author} {\bibfnamefont {V.~M.}\ \bibnamefont {Katkov}}, \
  and\ \bibinfo {author} {\bibfnamefont {V.~M.}\ \bibnamefont {Strakhovenko}},\
  }\href@noop {} {\emph {\bibinfo {title} {Electromagnetic Processes at High
  Energies in Oriented Single Crystals}}}\ (\bibinfo  {publisher} {World
  Scientific, Singapore},\ \bibinfo {year} {1994})\BibitemShut {NoStop}%
\bibitem [{\citenamefont {Berestetskii}\ \emph {et~al.}(1982)\citenamefont
  {Berestetskii}, , \citenamefont {Lifshitz},\ and\ \citenamefont
  {Pitevskii}}]{Landau_4}%
  \BibitemOpen
  \bibfield  {author} {\bibinfo {author} {\bibfnamefont {V.~B.}\ \bibnamefont
  {Berestetskii}}, , \bibinfo {author} {\bibfnamefont {E.~M.}\ \bibnamefont
  {Lifshitz}}, \ and\ \bibinfo {author} {\bibfnamefont {L.~P.}\ \bibnamefont
  {Pitevskii}},\ }\href@noop {} {\emph {\bibinfo {title} {Quantum
  electrodynamics}}}\ (\bibinfo  {publisher} {Pergamon, Oxford},\ \bibinfo
  {year} {1982})\BibitemShut {NoStop}%
\bibitem [{\citenamefont {Bulanov}\ \emph
  {et~al.}(2010{\natexlab{a}})\citenamefont {Bulanov}, \citenamefont {Mur},
  \citenamefont {Narozhny}, \citenamefont {Nees},\ and\ \citenamefont
  {Popov}}]{Bulanov_2010_a}%
  \BibitemOpen
  \bibfield  {author} {\bibinfo {author} {\bibfnamefont {S.~S.}\ \bibnamefont
  {Bulanov}}, \bibinfo {author} {\bibfnamefont {V.~D.}\ \bibnamefont {Mur}},
  \bibinfo {author} {\bibfnamefont {N.~B.}\ \bibnamefont {Narozhny}}, \bibinfo
  {author} {\bibfnamefont {J.}~\bibnamefont {Nees}}, \ and\ \bibinfo {author}
  {\bibfnamefont {V.~S.}\ \bibnamefont {Popov}},\ }\bibfield  {title} {\enquote
  {\bibinfo {title} {Multiple colliding electromagnetic pulses: A way to lower
  the threshold of $e^{+}e^{-}$ pair production from vacuum},}\ }\href@noop {}
  {\bibfield  {journal} {\bibinfo  {journal} {Phys. Rev. Lett.}\ }\textbf
  {\bibinfo {volume} {104}},\ \bibinfo {pages} {220404} (\bibinfo {year}
  {2010}{\natexlab{a}})}\BibitemShut {NoStop}%
\bibitem [{\citenamefont {Golla}\ \emph {et~al.}(2012)\citenamefont {Golla},
  \citenamefont {Chalopin}, \citenamefont {Bader}, \citenamefont {Harder},
  \citenamefont {Mantel}, \citenamefont {Maiwald}, \citenamefont {Lindlein},
  \citenamefont {Sondermann},\ and\ \citenamefont {Leuchs}}]{Golla_2012}%
  \BibitemOpen
  \bibfield  {author} {\bibinfo {author} {\bibfnamefont {Andrea}\ \bibnamefont
  {Golla}}, \bibinfo {author} {\bibfnamefont {Beno{\i}t}\ \bibnamefont
  {Chalopin}}, \bibinfo {author} {\bibfnamefont {Marianne}\ \bibnamefont
  {Bader}}, \bibinfo {author} {\bibfnamefont {Irina}\ \bibnamefont {Harder}},
  \bibinfo {author} {\bibfnamefont {Klaus}\ \bibnamefont {Mantel}}, \bibinfo
  {author} {\bibfnamefont {Robert}\ \bibnamefont {Maiwald}}, \bibinfo {author}
  {\bibfnamefont {Norbert}\ \bibnamefont {Lindlein}}, \bibinfo {author}
  {\bibfnamefont {Markus}\ \bibnamefont {Sondermann}}, \ and\ \bibinfo {author}
  {\bibfnamefont {Gerd}\ \bibnamefont {Leuchs}},\ }\bibfield  {title} {\enquote
  {\bibinfo {title} {Generation of a wave packet tailored to efficient free
  space excitation of a single atom},}\ }\href@noop {} {\bibfield  {journal}
  {\bibinfo  {journal} {The European Physical Journal D}\ }\textbf {\bibinfo
  {volume} {66}},\ \bibinfo {pages} {190} (\bibinfo {year} {2012})}\BibitemShut
  {NoStop}%
\bibitem [{\citenamefont {Gonoskov}\ \emph {et~al.}(2012)\citenamefont
  {Gonoskov}, \citenamefont {Aiello}, \citenamefont {Heugel},\ and\
  \citenamefont {Leuchs}}]{Gonoskov_2012}%
  \BibitemOpen
  \bibfield  {author} {\bibinfo {author} {\bibfnamefont {Ivan}\ \bibnamefont
  {Gonoskov}}, \bibinfo {author} {\bibfnamefont {Andrea}\ \bibnamefont
  {Aiello}}, \bibinfo {author} {\bibfnamefont {Simon}\ \bibnamefont {Heugel}},
  \ and\ \bibinfo {author} {\bibfnamefont {Gerd}\ \bibnamefont {Leuchs}},\
  }\bibfield  {title} {\enquote {\bibinfo {title} {Dipole pulse theory:
  Maximizing the field amplitude from $4\ensuremath{\pi}$ focused laser
  pulses},}\ }\href {\doibase 10.1103/PhysRevA.86.053836} {\bibfield  {journal}
  {\bibinfo  {journal} {Phys. Rev. A}\ }\textbf {\bibinfo {volume} {86}},\
  \bibinfo {pages} {053836} (\bibinfo {year} {2012})}\BibitemShut {NoStop}%
\bibitem [{\citenamefont {Bashinov}\ \emph {et~al.}(2013)\citenamefont
  {Bashinov}, \citenamefont {Gonoskov}, \citenamefont {Kim}, \citenamefont
  {Marklund}, \citenamefont {Mourou},\ and\ \citenamefont
  {Sergeev}}]{Bashinov_2013}%
  \BibitemOpen
  \bibfield  {author} {\bibinfo {author} {\bibfnamefont {Aleksei~V}\
  \bibnamefont {Bashinov}}, \bibinfo {author} {\bibfnamefont {Arkady~A}\
  \bibnamefont {Gonoskov}}, \bibinfo {author} {\bibfnamefont
  {Arkadii~Valentinovich}\ \bibnamefont {Kim}}, \bibinfo {author}
  {\bibfnamefont {Mattias}\ \bibnamefont {Marklund}}, \bibinfo {author}
  {\bibfnamefont {G{\'e}rard}\ \bibnamefont {Mourou}}, \ and\ \bibinfo {author}
  {\bibfnamefont {Aleksandr~M}\ \bibnamefont {Sergeev}},\ }\bibfield  {title}
  {\enquote {\bibinfo {title} {Electron acceleration and emission in a field of
  a plane and converging dipole wave of relativistic amplitudes with the
  radiation reaction force taken into account},}\ }\href@noop {} {\bibfield
  {journal} {\bibinfo  {journal} {Quantum Electronics}\ }\textbf {\bibinfo
  {volume} {43}},\ \bibinfo {pages} {291} (\bibinfo {year} {2013})}\BibitemShut
  {NoStop}%
\bibitem [{\citenamefont {Bashinov}\ \emph {et~al.}(2019)\citenamefont
  {Bashinov}, \citenamefont {Kumar},\ and\ \citenamefont
  {Efimenko}}]{Bashinov_2019}%
  \BibitemOpen
  \bibfield  {author} {\bibinfo {author} {\bibfnamefont {Aleksei~V}\
  \bibnamefont {Bashinov}}, \bibinfo {author} {\bibfnamefont {Punit}\
  \bibnamefont {Kumar}}, \ and\ \bibinfo {author} {\bibfnamefont
  {Evgenii~Sergeevich}\ \bibnamefont {Efimenko}},\ }\bibfield  {title}
  {\enquote {\bibinfo {title} {Confinement of electrons in the focus of the
  dipole wave},}\ }\href@noop {} {\bibfield  {journal} {\bibinfo  {journal}
  {Quantum Electronics}\ }\textbf {\bibinfo {volume} {49}},\ \bibinfo {pages}
  {314} (\bibinfo {year} {2019})}\BibitemShut {NoStop}%
\bibitem [{\citenamefont {Magnusson}\ \emph {et~al.}(2019)\citenamefont
  {Magnusson}, \citenamefont {Gonoskov}, \citenamefont {Marklund},
  \citenamefont {Esirkepov}, \citenamefont {Koga}, \citenamefont {Kondo},
  \citenamefont {Kando}, \citenamefont {Bulanov}, \citenamefont {Korn},
  \citenamefont {Geddes}, \citenamefont {Schroeder}, \citenamefont {Esarey},\
  and\ \citenamefont {Bulanov}}]{Magnusson_2019}%
  \BibitemOpen
  \bibfield  {author} {\bibinfo {author} {\bibfnamefont {J.}~\bibnamefont
  {Magnusson}}, \bibinfo {author} {\bibfnamefont {A.}~\bibnamefont {Gonoskov}},
  \bibinfo {author} {\bibfnamefont {M.}~\bibnamefont {Marklund}}, \bibinfo
  {author} {\bibfnamefont {T.~Zh.}\ \bibnamefont {Esirkepov}}, \bibinfo
  {author} {\bibfnamefont {J.~K.}\ \bibnamefont {Koga}}, \bibinfo {author}
  {\bibfnamefont {K.}~\bibnamefont {Kondo}}, \bibinfo {author} {\bibfnamefont
  {M.}~\bibnamefont {Kando}}, \bibinfo {author} {\bibfnamefont {S.~V.}\
  \bibnamefont {Bulanov}}, \bibinfo {author} {\bibfnamefont {G.}~\bibnamefont
  {Korn}}, \bibinfo {author} {\bibfnamefont {C.~G.~R.}\ \bibnamefont {Geddes}},
  \bibinfo {author} {\bibfnamefont {C.~B.}\ \bibnamefont {Schroeder}}, \bibinfo
  {author} {\bibfnamefont {E.}~\bibnamefont {Esarey}}, \ and\ \bibinfo {author}
  {\bibfnamefont {S.~S.}\ \bibnamefont {Bulanov}},\ }\bibfield  {title}
  {\enquote {\bibinfo {title} {Multiple colliding laser pulses as a basis for
  studying high-field high-energy physics},}\ }\href {\doibase
  10.1103/PhysRevA.100.063404} {\bibfield  {journal} {\bibinfo  {journal}
  {Phys. Rev. A}\ }\textbf {\bibinfo {volume} {100}},\ \bibinfo {pages}
  {063404} (\bibinfo {year} {2019})}\BibitemShut {NoStop}%
\bibitem [{\citenamefont {Kirk}\ \emph {et~al.}(2009)\citenamefont {Kirk},
  \citenamefont {Bell},\ and\ \citenamefont {Arka}}]{Kirk_2009}%
  \BibitemOpen
  \bibfield  {author} {\bibinfo {author} {\bibfnamefont {J~G}\ \bibnamefont
  {Kirk}}, \bibinfo {author} {\bibfnamefont {A~R}\ \bibnamefont {Bell}}, \ and\
  \bibinfo {author} {\bibfnamefont {I}~\bibnamefont {Arka}},\ }\bibfield
  {title} {\enquote {\bibinfo {title} {Pair production in counter-propagating
  laser beams},}\ }\href@noop {} {\bibfield  {journal} {\bibinfo  {journal}
  {Plasma Phys. Contr. F.}\ }\textbf {\bibinfo {volume} {51}},\ \bibinfo
  {pages} {085008} (\bibinfo {year} {2009})}\BibitemShut {NoStop}%
\bibitem [{\citenamefont {Bulanov}\ \emph
  {et~al.}(2010{\natexlab{b}})\citenamefont {Bulanov}, \citenamefont
  {Esirkepov}, \citenamefont {Thomas}, \citenamefont {Koga},\ and\
  \citenamefont {Bulanov}}]{Bulanov_2010}%
  \BibitemOpen
  \bibfield  {author} {\bibinfo {author} {\bibfnamefont {Stepan~S.}\
  \bibnamefont {Bulanov}}, \bibinfo {author} {\bibfnamefont {Timur~Zh.}\
  \bibnamefont {Esirkepov}}, \bibinfo {author} {\bibfnamefont {Alexander
  G.~R.}\ \bibnamefont {Thomas}}, \bibinfo {author} {\bibfnamefont {James~K.}\
  \bibnamefont {Koga}}, \ and\ \bibinfo {author} {\bibfnamefont {Sergei~V.}\
  \bibnamefont {Bulanov}},\ }\href@noop {} {\bibfield  {journal} {\bibinfo
  {journal} {Phys. Rev. Lett.}\ }\textbf {\bibinfo {volume} {105}},\ \bibinfo
  {pages} {220407} (\bibinfo {year} {2010}{\natexlab{b}})}\BibitemShut
  {NoStop}%
\bibitem [{\citenamefont {Gonoskov}\ \emph {et~al.}(2014)\citenamefont
  {Gonoskov}, \citenamefont {Bashinov}, \citenamefont {Gonoskov}, \citenamefont
  {Harvey}, \citenamefont {Ilderton}, \citenamefont {Kim}, \citenamefont
  {Marklund}, \citenamefont {Mourou},\ and\ \citenamefont
  {Sergeev}}]{Gonoskov_2014}%
  \BibitemOpen
  \bibfield  {author} {\bibinfo {author} {\bibfnamefont {A.}~\bibnamefont
  {Gonoskov}}, \bibinfo {author} {\bibfnamefont {A.}~\bibnamefont {Bashinov}},
  \bibinfo {author} {\bibfnamefont {I.}~\bibnamefont {Gonoskov}}, \bibinfo
  {author} {\bibfnamefont {C.}~\bibnamefont {Harvey}}, \bibinfo {author}
  {\bibfnamefont {A.}~\bibnamefont {Ilderton}}, \bibinfo {author}
  {\bibfnamefont {A.}~\bibnamefont {Kim}}, \bibinfo {author} {\bibfnamefont
  {M.}~\bibnamefont {Marklund}}, \bibinfo {author} {\bibfnamefont
  {G.}~\bibnamefont {Mourou}}, \ and\ \bibinfo {author} {\bibfnamefont
  {A.}~\bibnamefont {Sergeev}},\ }\bibfield  {title} {\enquote {\bibinfo
  {title} {Anomalous radiative trapping in laser fields of extreme
  intensity},}\ }\href@noop {} {\bibfield  {journal} {\bibinfo  {journal}
  {Phys. Rev. Lett.}\ }\textbf {\bibinfo {volume} {113}},\ \bibinfo {pages}
  {014801} (\bibinfo {year} {2014})}\BibitemShut {NoStop}%
\bibitem [{\citenamefont {Gong}\ \emph {et~al.}(2017)\citenamefont {Gong},
  \citenamefont {Hu}, \citenamefont {Shou}, \citenamefont {Qiao}, \citenamefont
  {Chen}, \citenamefont {He}, \citenamefont {Bulanov}, \citenamefont
  {Esirkepov}, \citenamefont {Bulanov},\ and\ \citenamefont {Yan}}]{Gong_2017}%
  \BibitemOpen
  \bibfield  {author} {\bibinfo {author} {\bibfnamefont {Z.}~\bibnamefont
  {Gong}}, \bibinfo {author} {\bibfnamefont {R.~H.}\ \bibnamefont {Hu}},
  \bibinfo {author} {\bibfnamefont {Y.~R.}\ \bibnamefont {Shou}}, \bibinfo
  {author} {\bibfnamefont {B.}~\bibnamefont {Qiao}}, \bibinfo {author}
  {\bibfnamefont {C.~E.}\ \bibnamefont {Chen}}, \bibinfo {author}
  {\bibfnamefont {X.~T.}\ \bibnamefont {He}}, \bibinfo {author} {\bibfnamefont
  {S.~S.}\ \bibnamefont {Bulanov}}, \bibinfo {author} {\bibfnamefont {T.~Zh.}\
  \bibnamefont {Esirkepov}}, \bibinfo {author} {\bibfnamefont {S.~V.}\
  \bibnamefont {Bulanov}}, \ and\ \bibinfo {author} {\bibfnamefont {X.~Q.}\
  \bibnamefont {Yan}},\ }\bibfield  {title} {\enquote {\bibinfo {title}
  {High-efficiency $\ensuremath{\gamma}$-ray flash generation via
  multiple-laser scattering in ponderomotive potential well},}\ }\href@noop {}
  {\bibfield  {journal} {\bibinfo  {journal} {Phys. Rev. E}\ }\textbf {\bibinfo
  {volume} {95}},\ \bibinfo {pages} {013210} (\bibinfo {year}
  {2017})}\BibitemShut {NoStop}%
\bibitem [{\citenamefont {Grismayer}\ \emph {et~al.}(2017)\citenamefont
  {Grismayer}, \citenamefont {Vranic}, \citenamefont {Martins}, \citenamefont
  {Fonseca},\ and\ \citenamefont {Silva}}]{Grismayer_2017}%
  \BibitemOpen
  \bibfield  {author} {\bibinfo {author} {\bibfnamefont {T.}~\bibnamefont
  {Grismayer}}, \bibinfo {author} {\bibfnamefont {M.}~\bibnamefont {Vranic}},
  \bibinfo {author} {\bibfnamefont {J.~L.}\ \bibnamefont {Martins}}, \bibinfo
  {author} {\bibfnamefont {R.~A.}\ \bibnamefont {Fonseca}}, \ and\ \bibinfo
  {author} {\bibfnamefont {L.~O.}\ \bibnamefont {Silva}},\ }\bibfield  {title}
  {\enquote {\bibinfo {title} {Seeded qed cascades in counterpropagating laser
  pulses},}\ }\href {\doibase 10.1103/PhysRevE.95.023210} {\bibfield  {journal}
  {\bibinfo  {journal} {Phys. Rev. E}\ }\textbf {\bibinfo {volume} {95}},\
  \bibinfo {pages} {023210} (\bibinfo {year} {2017})}\BibitemShut {NoStop}%
\bibitem [{\citenamefont {Grismayer}\ \emph {et~al.}(2016)\citenamefont
  {Grismayer}, \citenamefont {Vranic}, \citenamefont {Martins}, \citenamefont
  {Fonseca},\ and\ \citenamefont {Silva}}]{Grismayer_2016}%
  \BibitemOpen
  \bibfield  {author} {\bibinfo {author} {\bibfnamefont {T.}~\bibnamefont
  {Grismayer}}, \bibinfo {author} {\bibfnamefont {M.}~\bibnamefont {Vranic}},
  \bibinfo {author} {\bibfnamefont {J.~L.}\ \bibnamefont {Martins}}, \bibinfo
  {author} {\bibfnamefont {R.~A.}\ \bibnamefont {Fonseca}}, \ and\ \bibinfo
  {author} {\bibfnamefont {L.~O.}\ \bibnamefont {Silva}},\ }\bibfield  {title}
  {\enquote {\bibinfo {title} {Laser absorption via quantum electrodynamics
  cascades in counter propagating laser pulses},}\ }\href@noop {} {\bibfield
  {journal} {\bibinfo  {journal} {Phys. Plasmas}\ }\textbf {\bibinfo {volume}
  {23}},\ \bibinfo {pages} {056706} (\bibinfo {year} {2016})}\BibitemShut
  {NoStop}%
\bibitem [{\citenamefont {Jirka}\ \emph {et~al.}(2016)\citenamefont {Jirka},
  \citenamefont {Klimo}, \citenamefont {Bulanov}, \citenamefont {Esirkepov},
  \citenamefont {Gelfer}, \citenamefont {Bulanov}, \citenamefont {Weber},\ and\
  \citenamefont {Korn}}]{Jirka_2016}%
  \BibitemOpen
  \bibfield  {author} {\bibinfo {author} {\bibfnamefont {M.}~\bibnamefont
  {Jirka}}, \bibinfo {author} {\bibfnamefont {O.}~\bibnamefont {Klimo}},
  \bibinfo {author} {\bibfnamefont {S.~V.}\ \bibnamefont {Bulanov}}, \bibinfo
  {author} {\bibfnamefont {T.~Zh.}\ \bibnamefont {Esirkepov}}, \bibinfo
  {author} {\bibfnamefont {E.}~\bibnamefont {Gelfer}}, \bibinfo {author}
  {\bibfnamefont {S.~S.}\ \bibnamefont {Bulanov}}, \bibinfo {author}
  {\bibfnamefont {S.}~\bibnamefont {Weber}}, \ and\ \bibinfo {author}
  {\bibfnamefont {G.}~\bibnamefont {Korn}},\ }\bibfield  {title} {\enquote
  {\bibinfo {title} {Electron dynamics and $\ensuremath{\gamma}$ and
  ${e}^{\ensuremath{-}}{e}^{+}$ production by colliding laser pulses},}\
  }\href@noop {} {\bibfield  {journal} {\bibinfo  {journal} {Phys. Rev. E}\
  }\textbf {\bibinfo {volume} {93}},\ \bibinfo {pages} {023207} (\bibinfo
  {year} {2016})}\BibitemShut {NoStop}%
\bibitem [{\citenamefont {Lv}\ \emph {et~al.}(2022)\citenamefont {Lv},
  \citenamefont {Raicher}, \citenamefont {Keitel},\ and\ \citenamefont
  {Hatsagortsyan}}]{Lv_2022}%
  \BibitemOpen
  \bibfield  {author} {\bibinfo {author} {\bibfnamefont {QZ}~\bibnamefont
  {Lv}}, \bibinfo {author} {\bibfnamefont {E}~\bibnamefont {Raicher}}, \bibinfo
  {author} {\bibfnamefont {CH}~\bibnamefont {Keitel}}, \ and\ \bibinfo {author}
  {\bibfnamefont {KZ}~\bibnamefont {Hatsagortsyan}},\ }\bibfield  {title}
  {\enquote {\bibinfo {title} {High-brilliance ultranarrow-band x rays via
  electron radiation in colliding laser pulses},}\ }\href@noop {} {\bibfield
  {journal} {\bibinfo  {journal} {Physical Review Letters}\ }\textbf {\bibinfo
  {volume} {128}},\ \bibinfo {pages} {024801} (\bibinfo {year}
  {2022})}\BibitemShut {NoStop}%
\bibitem [{\citenamefont {Brezin}\ and\ \citenamefont
  {Itzykson}(1970)}]{Brezin_1970}%
  \BibitemOpen
  \bibfield  {author} {\bibinfo {author} {\bibfnamefont {E.}~\bibnamefont
  {Brezin}}\ and\ \bibinfo {author} {\bibfnamefont {C.}~\bibnamefont
  {Itzykson}},\ }\bibfield  {title} {\enquote {\bibinfo {title} {Pair
  production in vacuum by an alternating field},}\ }\href@noop {} {\bibfield
  {journal} {\bibinfo  {journal} {Phys. Rev. D}\ }\textbf {\bibinfo {volume}
  {2}},\ \bibinfo {pages} {1191--1199} (\bibinfo {year} {1970})}\BibitemShut
  {NoStop}%
\bibitem [{\citenamefont {Raicher}\ and\ \citenamefont
  {Hatsagortsyan}(2020)}]{Raicher_2020}%
  \BibitemOpen
  \bibfield  {author} {\bibinfo {author} {\bibfnamefont {E.}~\bibnamefont
  {Raicher}}\ and\ \bibinfo {author} {\bibfnamefont {K.~Z.}\ \bibnamefont
  {Hatsagortsyan}},\ }\bibfield  {title} {\enquote {\bibinfo {title} {Nonlinear
  qed in an ultrastrong rotating electric field: Signatures of the
  momentum-dependent effective mass},}\ }\href {\doibase
  10.1103/PhysRevResearch.2.013240} {\bibfield  {journal} {\bibinfo  {journal}
  {Phys. Rev. Research}\ }\textbf {\bibinfo {volume} {2}},\ \bibinfo {pages}
  {013240} (\bibinfo {year} {2020})}\BibitemShut {NoStop}%
\bibitem [{\citenamefont {Villalba-Ch\'avez}\ and\ \citenamefont
  {M\"uller}(2019)}]{Villalba_2019}%
  \BibitemOpen
  \bibfield  {author} {\bibinfo {author} {\bibfnamefont {Selym}\ \bibnamefont
  {Villalba-Ch\'avez}}\ and\ \bibinfo {author} {\bibfnamefont {Carsten}\
  \bibnamefont {M\"uller}},\ }\bibfield  {title} {\enquote {\bibinfo {title}
  {Signatures of the schwinger mechanism assisted by a fast-oscillating
  electric field},}\ }\href {\doibase 10.1103/PhysRevD.100.116018} {\bibfield
  {journal} {\bibinfo  {journal} {Phys. Rev. D}\ }\textbf {\bibinfo {volume}
  {100}},\ \bibinfo {pages} {116018} (\bibinfo {year} {2019})}\BibitemShut
  {NoStop}%
\bibitem [{\citenamefont {Sch\"utzhold}\ \emph {et~al.}(2008)\citenamefont
  {Sch\"utzhold}, \citenamefont {Gies},\ and\ \citenamefont
  {Dunne}}]{Dunne_2008}%
  \BibitemOpen
  \bibfield  {author} {\bibinfo {author} {\bibfnamefont {Ralf}\ \bibnamefont
  {Sch\"utzhold}}, \bibinfo {author} {\bibfnamefont {Holger}\ \bibnamefont
  {Gies}}, \ and\ \bibinfo {author} {\bibfnamefont {Gerald}\ \bibnamefont
  {Dunne}},\ }\bibfield  {title} {\enquote {\bibinfo {title} {Dynamically
  assisted schwinger mechanism},}\ }\href {\doibase
  10.1103/PhysRevLett.101.130404} {\bibfield  {journal} {\bibinfo  {journal}
  {Phys. Rev. Lett.}\ }\textbf {\bibinfo {volume} {101}},\ \bibinfo {pages}
  {130404} (\bibinfo {year} {2008})}\BibitemShut {NoStop}%
\bibitem [{\citenamefont {Kirk}(2016)}]{Kirk_2016}%
  \BibitemOpen
  \bibfield  {author} {\bibinfo {author} {\bibfnamefont {J~G}\ \bibnamefont
  {Kirk}},\ }\bibfield  {title} {\enquote {\bibinfo {title} {Radiative trapping
  in intense laser beams},}\ }\href@noop {} {\bibfield  {journal} {\bibinfo
  {journal} {Plasma Phys. Cont. Fus.}\ }\textbf {\bibinfo {volume} {58}},\
  \bibinfo {pages} {085005} (\bibinfo {year} {2016})}\BibitemShut {NoStop}%
\bibitem [{\citenamefont {Kapitza}\ and\ \citenamefont
  {Dirac}(1933)}]{Kapitza_1933}%
  \BibitemOpen
  \bibfield  {author} {\bibinfo {author} {\bibfnamefont {P.~L.}\ \bibnamefont
  {Kapitza}}\ and\ \bibinfo {author} {\bibfnamefont {P.~A.~M.}\ \bibnamefont
  {Dirac}},\ }\bibfield  {title} {\enquote {\bibinfo {title} {The reflection of
  electrons from standing light waves},}\ }\href@noop {} {\bibfield  {journal}
  {\bibinfo  {journal} {Math. Proc. Cambr. Phil. Soc.}\ }\textbf {\bibinfo
  {volume} {29}},\ \bibinfo {pages} {297–300} (\bibinfo {year}
  {1933})}\BibitemShut {NoStop}%
\bibitem [{\citenamefont {Batelaan}(2007)}]{Batelaan_2007}%
  \BibitemOpen
  \bibfield  {author} {\bibinfo {author} {\bibfnamefont {H.}~\bibnamefont
  {Batelaan}},\ }\bibfield  {title} {\enquote {\bibinfo {title} {Colloquium:
  Illuminating the kapitza-dirac effect with electron matter optics},}\
  }\href@noop {} {\bibfield  {journal} {\bibinfo  {journal} {Rev. Mod. Phys.}\
  }\textbf {\bibinfo {volume} {79}},\ \bibinfo {pages} {929--941} (\bibinfo
  {year} {2007})}\BibitemShut {NoStop}%
\bibitem [{\citenamefont {Ahrens}\ \emph {et~al.}(2012)\citenamefont {Ahrens},
  \citenamefont {Bauke}, \citenamefont {Keitel},\ and\ \citenamefont
  {M\"uller}}]{Ahrens_2012}%
  \BibitemOpen
  \bibfield  {author} {\bibinfo {author} {\bibfnamefont {Sven}\ \bibnamefont
  {Ahrens}}, \bibinfo {author} {\bibfnamefont {Heiko}\ \bibnamefont {Bauke}},
  \bibinfo {author} {\bibfnamefont {Christoph~H.}\ \bibnamefont {Keitel}}, \
  and\ \bibinfo {author} {\bibfnamefont {Carsten}\ \bibnamefont {M\"uller}},\
  }\bibfield  {title} {\enquote {\bibinfo {title} {Spin dynamics in the
  kapitza-dirac effect},}\ }\href@noop {} {\bibfield  {journal} {\bibinfo
  {journal} {Phys. Rev. Lett.}\ }\textbf {\bibinfo {volume} {109}},\ \bibinfo
  {pages} {043601} (\bibinfo {year} {2012})}\BibitemShut {NoStop}%
\bibitem [{\citenamefont {Dellweg}\ and\ \citenamefont
  {M\"uller}(2017)}]{Mueller_2017}%
  \BibitemOpen
  \bibfield  {author} {\bibinfo {author} {\bibfnamefont {M.M.}\ \bibnamefont
  {Dellweg}}\ and\ \bibinfo {author} {\bibfnamefont {C.}~\bibnamefont
  {M\"uller}},\ }\bibfield  {title} {\enquote {\bibinfo {title}
  {Spin-polarizing interferometric beam splitter for free electrons},}\
  }\href@noop {} {\bibfield  {journal} {\bibinfo  {journal} {Phys. Rev. Lett.}\
  }\textbf {\bibinfo {volume} {118}},\ \bibinfo {pages} {070403} (\bibinfo
  {year} {2017})}\BibitemShut {NoStop}%
\bibitem [{\citenamefont {Friedman}\ \emph {et~al.}(1988)\citenamefont
  {Friedman}, \citenamefont {Gover}, \citenamefont {Kurizki}, \citenamefont
  {Ruschin},\ and\ \citenamefont {Yariv}}]{Friedman_1988}%
  \BibitemOpen
  \bibfield  {author} {\bibinfo {author} {\bibfnamefont {A.}~\bibnamefont
  {Friedman}}, \bibinfo {author} {\bibfnamefont {A.}~\bibnamefont {Gover}},
  \bibinfo {author} {\bibfnamefont {G.}~\bibnamefont {Kurizki}}, \bibinfo
  {author} {\bibfnamefont {S.}~\bibnamefont {Ruschin}}, \ and\ \bibinfo
  {author} {\bibfnamefont {A.}~\bibnamefont {Yariv}},\ }\bibfield  {title}
  {\enquote {\bibinfo {title} {Spontaneous and stimulated emission from
  quasifree electrons},}\ }\href@noop {} {\bibfield  {journal} {\bibinfo
  {journal} {Rev. Mod. Phys.}\ }\textbf {\bibinfo {volume} {60}},\ \bibinfo
  {pages} {471--535} (\bibinfo {year} {1988})}\BibitemShut {NoStop}%
\bibitem [{\citenamefont {{Pantell}}\ \emph {et~al.}(1968)\citenamefont
  {{Pantell}}, \citenamefont {{Soncini}},\ and\ \citenamefont
  {{Puthoff}}}]{Pantell_1968}%
  \BibitemOpen
  \bibfield  {author} {\bibinfo {author} {\bibfnamefont {R.}~\bibnamefont
  {{Pantell}}}, \bibinfo {author} {\bibfnamefont {G.}~\bibnamefont
  {{Soncini}}}, \ and\ \bibinfo {author} {\bibfnamefont {H.}~\bibnamefont
  {{Puthoff}}},\ }\bibfield  {title} {\enquote {\bibinfo {title} {Stimulated
  photon-electron scattering},}\ }\href@noop {} {\bibfield  {journal} {\bibinfo
   {journal} {IEEE J. Quant. El.}\ }\textbf {\bibinfo {volume} {4}},\ \bibinfo
  {pages} {905--907} (\bibinfo {year} {1968})}\BibitemShut {NoStop}%
\bibitem [{\citenamefont {Fedorov}(1981)}]{Fedorov_1981}%
  \BibitemOpen
  \bibfield  {author} {\bibinfo {author} {\bibfnamefont {M.V.}\ \bibnamefont
  {Fedorov}},\ }\bibfield  {title} {\enquote {\bibinfo {title} {Free-electron
  lasers and multiphoton free-free transitions},}\ }\href@noop {} {\bibfield
  {journal} {\bibinfo  {journal} {Progress in Quantum Electronics}\ }\textbf
  {\bibinfo {volume} {7}},\ \bibinfo {pages} {73 -- 116} (\bibinfo {year}
  {1981})}\BibitemShut {NoStop}%
\bibitem [{\citenamefont {Avetissian}(2016)}]{Avetissian_b_2016}%
  \BibitemOpen
  \bibfield  {author} {\bibinfo {author} {\bibfnamefont {H.~K.}\ \bibnamefont
  {Avetissian}},\ }\href@noop {} {\emph {\bibinfo {title} {Relativistic
  nonlinear electrodynamics}}}\ (\bibinfo  {publisher} {Springer, New York},\
  \bibinfo {year} {2016})\BibitemShut {NoStop}%
\bibitem [{\citenamefont {Saldin}\ \emph {et~al.}(1995)\citenamefont {Saldin},
  \citenamefont {Schneidmiller},\ and\ \citenamefont {Yurkov}}]{Saldin_1995}%
  \BibitemOpen
  \bibfield  {author} {\bibinfo {author} {\bibfnamefont {E.~L.}\ \bibnamefont
  {Saldin}}, \bibinfo {author} {\bibfnamefont {E.~A.}\ \bibnamefont
  {Schneidmiller}}, \ and\ \bibinfo {author} {\bibfnamefont {M.~V.}\
  \bibnamefont {Yurkov}},\ }\bibfield  {title} {\enquote {\bibinfo {title} {The
  physics of free electron lasers. an introduction},}\ }\href@noop {}
  {\bibfield  {journal} {\bibinfo  {journal} {Phys. Rep.}\ }\textbf {\bibinfo
  {volume} {260}},\ \bibinfo {pages} {187 -- 327} (\bibinfo {year}
  {1995})}\BibitemShut {NoStop}%
\bibitem [{\citenamefont {King}\ and\ \citenamefont {Hu}(2016)}]{King_2016}%
  \BibitemOpen
  \bibfield  {author} {\bibinfo {author} {\bibfnamefont {B.}~\bibnamefont
  {King}}\ and\ \bibinfo {author} {\bibfnamefont {H.}~\bibnamefont {Hu}},\
  }\bibfield  {title} {\enquote {\bibinfo {title} {Classical and quantum
  dynamics of a charged scalar particle in a background of two
  counterpropagating plane waves},}\ }\href@noop {} {\bibfield  {journal}
  {\bibinfo  {journal} {Phys. Rev. D}\ }\textbf {\bibinfo {volume} {94}},\
  \bibinfo {pages} {125010} (\bibinfo {year} {2016})}\BibitemShut {NoStop}%
\bibitem [{\citenamefont {Hu}\ and\ \citenamefont {Huang}(2015)}]{Hu_2015}%
  \BibitemOpen
  \bibfield  {author} {\bibinfo {author} {\bibfnamefont {H.}~\bibnamefont
  {Hu}}\ and\ \bibinfo {author} {\bibfnamefont {J.}~\bibnamefont {Huang}},\
  }\bibfield  {title} {\enquote {\bibinfo {title} {Analytical solution for the
  klein-gordon equation and action function of the solution for the dirac
  equation in counterpropagating laser waves},}\ }\href@noop {} {\bibfield
  {journal} {\bibinfo  {journal} {Phys. Rev. A}\ }\textbf {\bibinfo {volume}
  {92}},\ \bibinfo {pages} {062105} (\bibinfo {year} {2015})}\BibitemShut
  {NoStop}%
\bibitem [{\citenamefont {Lv}\ \emph {et~al.}(2021{\natexlab{b}})\citenamefont
  {Lv}, \citenamefont {Raicher}, \citenamefont {Keitel},\ and\ \citenamefont
  {Hatsagortsyan}}]{Lv_2021b}%
  \BibitemOpen
  \bibfield  {author} {\bibinfo {author} {\bibfnamefont {Q~Z}\ \bibnamefont
  {Lv}}, \bibinfo {author} {\bibfnamefont {E}~\bibnamefont {Raicher}}, \bibinfo
  {author} {\bibfnamefont {C~H}\ \bibnamefont {Keitel}}, \ and\ \bibinfo
  {author} {\bibfnamefont {K~Z}\ \bibnamefont {Hatsagortsyan}},\ }\bibfield
  {title} {\enquote {\bibinfo {title} {Ultrarelativistic electrons in
  counterpropagating laser beams},}\ }\href {\doibase 10.1088/1367-2630/abfa60}
  {\bibfield  {journal} {\bibinfo  {journal} {New Journal of Physics}\ }\textbf
  {\bibinfo {volume} {23}},\ \bibinfo {pages} {065005} (\bibinfo {year}
  {2021}{\natexlab{b}})}\BibitemShut {NoStop}%
\bibitem [{\citenamefont {Ritus}(1985)}]{Ritus_1985}%
  \BibitemOpen
  \bibfield  {author} {\bibinfo {author} {\bibfnamefont {V.~I.}\ \bibnamefont
  {Ritus}},\ }\href@noop {} {\bibfield  {journal} {\bibinfo  {journal} {J. Sov.
  Laser Res.}\ }\textbf {\bibinfo {volume} {6}},\ \bibinfo {pages} {497}
  (\bibinfo {year} {1985})}\BibitemShut {NoStop}%
\bibitem [{\citenamefont {Otto}\ \emph {et~al.}(2015)\citenamefont {Otto},
  \citenamefont {Seipt}, \citenamefont {Blaschke}, \citenamefont {Smolyansky},\
  and\ \citenamefont {K{\"a}mpfer}}]{Otto_2015}%
  \BibitemOpen
  \bibfield  {author} {\bibinfo {author} {\bibfnamefont {Andreas}\ \bibnamefont
  {Otto}}, \bibinfo {author} {\bibfnamefont {Daniel}\ \bibnamefont {Seipt}},
  \bibinfo {author} {\bibfnamefont {David}\ \bibnamefont {Blaschke}}, \bibinfo
  {author} {\bibfnamefont {Stanislav~Alexandrovich}\ \bibnamefont
  {Smolyansky}}, \ and\ \bibinfo {author} {\bibfnamefont {Burkhard}\
  \bibnamefont {K{\"a}mpfer}},\ }\bibfield  {title} {\enquote {\bibinfo {title}
  {Dynamical schwinger process in a bifrequent electric field of finite
  duration: survey on amplification},}\ }\href@noop {} {\bibfield  {journal}
  {\bibinfo  {journal} {Physical Review D}\ }\textbf {\bibinfo {volume} {91}},\
  \bibinfo {pages} {105018} (\bibinfo {year} {2015})}\BibitemShut {NoStop}%
\bibitem [{\citenamefont {Schneider}\ and\ \citenamefont
  {Sch{\"u}tzhold}(2016)}]{Schneider_2016}%
  \BibitemOpen
  \bibfield  {author} {\bibinfo {author} {\bibfnamefont {Christian}\
  \bibnamefont {Schneider}}\ and\ \bibinfo {author} {\bibfnamefont {Ralf}\
  \bibnamefont {Sch{\"u}tzhold}},\ }\bibfield  {title} {\enquote {\bibinfo
  {title} {Dynamically assisted sauter-schwinger effect in inhomogeneous
  electric fields},}\ }\href@noop {} {\bibfield  {journal} {\bibinfo  {journal}
  {Journal of High Energy Physics}\ }\textbf {\bibinfo {volume} {2016}},\
  \bibinfo {pages} {164} (\bibinfo {year} {2016})}\BibitemShut {NoStop}%
\bibitem [{\citenamefont {Torgrimsson}\ \emph {et~al.}(2017)\citenamefont
  {Torgrimsson}, \citenamefont {Schneider}, \citenamefont {Oertel},\ and\
  \citenamefont {Sch{\"u}tzhold}}]{Torgrimsson_2017}%
  \BibitemOpen
  \bibfield  {author} {\bibinfo {author} {\bibfnamefont {Greger}\ \bibnamefont
  {Torgrimsson}}, \bibinfo {author} {\bibfnamefont {Christian}\ \bibnamefont
  {Schneider}}, \bibinfo {author} {\bibfnamefont {Johannes}\ \bibnamefont
  {Oertel}}, \ and\ \bibinfo {author} {\bibfnamefont {Ralf}\ \bibnamefont
  {Sch{\"u}tzhold}},\ }\bibfield  {title} {\enquote {\bibinfo {title}
  {Dynamically assisted sauter-schwinger effect—non-perturbative versus
  perturbative aspects},}\ }\href@noop {} {\bibfield  {journal} {\bibinfo
  {journal} {Journal of High Energy Physics}\ }\textbf {\bibinfo {volume}
  {2017}},\ \bibinfo {pages} {43} (\bibinfo {year} {2017})}\BibitemShut
  {NoStop}%
\bibitem [{\citenamefont {Torgrimsson}\ \emph {et~al.}(2018)\citenamefont
  {Torgrimsson}, \citenamefont {Schneider},\ and\ \citenamefont
  {Sch{\"u}tzhold}}]{Torgrimsson_2018}%
  \BibitemOpen
  \bibfield  {author} {\bibinfo {author} {\bibfnamefont {Greger}\ \bibnamefont
  {Torgrimsson}}, \bibinfo {author} {\bibfnamefont {Christian}\ \bibnamefont
  {Schneider}}, \ and\ \bibinfo {author} {\bibfnamefont {Ralf}\ \bibnamefont
  {Sch{\"u}tzhold}},\ }\bibfield  {title} {\enquote {\bibinfo {title}
  {Sauter-schwinger pair creation dynamically assisted by a plane wave},}\
  }\href@noop {} {\bibfield  {journal} {\bibinfo  {journal} {Physical Review
  D}\ }\textbf {\bibinfo {volume} {97}},\ \bibinfo {pages} {096004} (\bibinfo
  {year} {2018})}\BibitemShut {NoStop}%
\bibitem [{\citenamefont {Linder}\ \emph {et~al.}(2015)\citenamefont {Linder},
  \citenamefont {Schneider}, \citenamefont {Sicking}, \citenamefont {Szpak},\
  and\ \citenamefont {Sch{\"u}tzhold}}]{Linder_2015}%
  \BibitemOpen
  \bibfield  {author} {\bibinfo {author} {\bibfnamefont {Malte~F}\ \bibnamefont
  {Linder}}, \bibinfo {author} {\bibfnamefont {Christian}\ \bibnamefont
  {Schneider}}, \bibinfo {author} {\bibfnamefont {Joachim}\ \bibnamefont
  {Sicking}}, \bibinfo {author} {\bibfnamefont {Nikodem}\ \bibnamefont
  {Szpak}}, \ and\ \bibinfo {author} {\bibfnamefont {Ralf}\ \bibnamefont
  {Sch{\"u}tzhold}},\ }\bibfield  {title} {\enquote {\bibinfo {title} {Pulse
  shape dependence in the dynamically assisted sauter-schwinger effect},}\
  }\href@noop {} {\bibfield  {journal} {\bibinfo  {journal} {Physical Review
  D}\ }\textbf {\bibinfo {volume} {92}},\ \bibinfo {pages} {085009} (\bibinfo
  {year} {2015})}\BibitemShut {NoStop}%
\bibitem [{\citenamefont {Aleksandrov}\ \emph {et~al.}(2018)\citenamefont
  {Aleksandrov}, \citenamefont {Plunien},\ and\ \citenamefont
  {Shabaev}}]{Aleksandrov_2018}%
  \BibitemOpen
  \bibfield  {author} {\bibinfo {author} {\bibfnamefont {IA}~\bibnamefont
  {Aleksandrov}}, \bibinfo {author} {\bibfnamefont {G}~\bibnamefont {Plunien}},
  \ and\ \bibinfo {author} {\bibfnamefont {VM}~\bibnamefont {Shabaev}},\
  }\bibfield  {title} {\enquote {\bibinfo {title} {Dynamically assisted
  schwinger effect beyond the spatially-uniform-field approximation},}\
  }\href@noop {} {\bibfield  {journal} {\bibinfo  {journal} {Physical Review
  D}\ }\textbf {\bibinfo {volume} {97}},\ \bibinfo {pages} {116001} (\bibinfo
  {year} {2018})}\BibitemShut {NoStop}%
\bibitem [{\citenamefont {Di~Piazza}\ \emph {et~al.}(2009)\citenamefont
  {Di~Piazza}, \citenamefont {L{\"o}tstedt}, \citenamefont {Milstein},\ and\
  \citenamefont {Keitel}}]{DiPiazza_2009}%
  \BibitemOpen
  \bibfield  {author} {\bibinfo {author} {\bibfnamefont {A}~\bibnamefont
  {Di~Piazza}}, \bibinfo {author} {\bibfnamefont {E}~\bibnamefont
  {L{\"o}tstedt}}, \bibinfo {author} {\bibfnamefont {AI}~\bibnamefont
  {Milstein}}, \ and\ \bibinfo {author} {\bibfnamefont {CH}~\bibnamefont
  {Keitel}},\ }\bibfield  {title} {\enquote {\bibinfo {title} {Barrier control
  in tunneling $e^+ - e^-$ photoproduction},}\ }\href@noop {} {\bibfield
  {journal} {\bibinfo  {journal} {Physical Review Letters}\ }\textbf {\bibinfo
  {volume} {103}},\ \bibinfo {pages} {170403} (\bibinfo {year}
  {2009})}\BibitemShut {NoStop}%
\end{thebibliography}%

\end{document}